\documentclass[reprint,twocolumn,aps,amsmath,amssymb,prl,superscriptaddress,floatfix,tightenlines]{revtex4-2}
\usepackage{amsmath, amssymb, amsthm}
\usepackage{mathtools}
\usepackage{color}
\usepackage{braket}
\usepackage{comment}
\usepackage{bbm}
\usepackage{bm}
\usepackage{hyperref}
\usepackage{url}
\usepackage[dvipsnames]{xcolor}
\usepackage{cleveref}
\usepackage{times,txfonts}
\usepackage[T1]{fontenc}

\hypersetup{
    bookmarksnumbered=true,
    unicode=false,
    pdfstartview={FitH},
    pdftitle={},
    pdfauthor={},
    pdfsubject={},
    pdfcreator={},
    pdfproducer={},
    pdfkeywords={},
    pdfnewwindow=true,
    colorlinks=true,
    linkcolor=NavyBlue,
    citecolor=NavyBlue,
    filecolor=NavyBlue,
    urlcolor=NavyBlue
}

\DeclareMathOperator{\tr}{Tr}

\allowdisplaybreaks[4]

\theoremstyle{plain}
\newtheorem{theorem}{Theorem}

\theoremstyle{definition}
\newtheorem{definition}[theorem]{Definition}

\theoremstyle{remark}

\newcommand{\rt}[1]{{#1}}

\begin{document}
\author{Kohdai Kuroiwa}
\affiliation{Institute for Quantum Computing and Department of Combinatorics and Optimization, University of Waterloo, Ontario, Canada, N2L 3G1}
\affiliation{Perimeter Institute for Theoretical Physics, Ontario, Canada, N2L 2Y5}
\author{Ryuji Takagi}
\affiliation{Department of Basic Science, The University of Tokyo, 3-8-1 Komaba, Meguro-ku, Tokyo, 153-8902, Japan}
\author{Gerardo Adesso}
\affiliation{School of Mathematical Sciences and Centre for the Mathematical and Theoretical Physics of Quantum Non-Equilibrium Systems, University of Nottingham, University Park, Nottingham, NG7 2RD, United Kingdom}
\author{Hayata Yamasaki}
\affiliation{Department of Physics, Graduate School of Science, The University of Tokyo, 7-3-1 Hongo, Bunkyo-ku, Tokyo, 113-0033, Japan}
\title{Every quantum helps: 
Operational advantage of quantum resources beyond convexity}

\begin{abstract}
    Identifying what quantum-mechanical properties are useful to untap a superior performance in quantum technologies is a pivotal question. 
    Quantum resource theories provide a unified framework to analyze and understand such properties, as successfully demonstrated for entanglement and coherence. While these are examples of convex resources, for which quantum advantages can always be identified,  many physical resources are described by a non-convex set of free states and their interpretation has so far remained elusive. Here we address the fundamental question of the usefulness of quantum resources without convexity assumption, by providing two operational interpretations of the generalized robustness measure in general resource theories.   
    First, we characterize the generalized robustness in terms of a non-linear resource witness and reveal that any state is more advantageous than a free one in some multi-copy channel discrimination task.
    Next, we consider a scenario where a theory is characterized by multiple constraints and show that the generalized robustness coincides with the worst-case advantage in a single-copy channel discrimination setting.
    Based on these characterizations, we conclude that every quantum resource state shows a qualitative and quantitative advantage in discrimination problems in a general resource theory even without any specification on the structure of the free states. 
\end{abstract}

\maketitle

\textbf{\textit{Introduction}}.---
The drive to interpret and characterize signature properties of quantum mechanics spurred the rise of quantum information theory and shed new light on foundational questions of modern science \cite{Nobel2022}. The further realization that such properties can be exploited as {\it resources} to enhance computational power, sensing precision, communication security, and many more tasks unlocked a technological overhaul whose impact is unfolding at an astounding pace \cite{Dowling2003}. 
Under the hood of such impressive advances, key questions remain: What fundamental ingredients are needed for the optimal performance of quantum technologies? Does {\it every} quantum feature lead to an advantage in practical applications? 

Addressing these and related questions can be facilitated by the study of {\it quantum resource theories} (QRTs)~\cite{Streltsov2017,Chitambar2018,Kuroiwa2020}, which classify quantum states and operations into free (not useful, akin to `classical') versus non-free (potentially useful as `resources'), and provide a mathematically rigorous formalism to validate various measures of quantum properties and investigate their operational significance. 
The \textit{generalized robustness}~\cite{Vidal1999,Steiner2003,Harrow2003} is one such prominent quantifier in the framework of QRTs. It is calculable with semidefinite programming for several representative cases, including  $k$-entanglement~\cite{Bae2019}, coherence~\cite{Napoli2016}, multilevel coherence~\cite{Ringbauer2018}, asymmetry~\cite{Piani2016}, magic~\cite{Liu_ZW2019a,Seddon2021}, and steering~\cite{Piani2015}. 
It also has an operational characterization, which was initially shown for steering~\cite{Piani2015}, coherence~\cite{Napoli2016}, and asymmetry~\cite{Piani2016}. 
Notably, Ref.~\cite{Takagi2019b} generalized these results showing that, in any {\it convex} QRT on finite-dimensional state spaces, the generalized robustness can be interpreted operationally as the advantage of a quantum resource state in some channel discrimination task~\cite{Kitaev_1997,Andrew2000,Acin2001,watrous_2018}. This result was recently extended to dynamical resource theories~\cite{Takagi2019a,Uola2019}, general probabilistic theories~\cite{Takagi2019a}, and infinite-dimensional convex QRTs~\cite{Regula2021,Lami2021}. However, these analyses heavily rely on the convex geometry of the set of resource-free states, which guarantees the connection between the generalized robustness and \textit{resource witness}~\cite{Brandao2005,Eisert_2007}. A straightforward generalization of these operational approaches is no longer applicable when the convexity assumption is dropped. 

Nevertheless, physically well-motivated quantum resources do not necessarily have a convex structure.
For example, non-Gaussianity is widely regarded as an advantageous or even essential resource for quantum optical technologies \cite{Mattia2021}, yet the set of Gaussian states in continuous variable systems~\cite{Weedbrook2012,Adesso2014Gauss} is non-convex. The set of quantum Markov chains~\cite{Wakakuwa2017}, as well as the sets of states with no quantum discord \cite{Ollivier2001,Henderson2001}, e.g., classical-classical and classical-quantum states~\cite{ABC2016,Bera2017}, are also not convex. Furthermore, tailored resource theories in which physical limitations are identified from experimental constraints are not a priori expected to be built upon a convex set of free states. In fact, classical randomness can be regarded as a resource~\cite{Groisman2005,Anshu2018}, meaning that convexity itself may well be expensive in general.  
Hence, the current lack of techniques towards resources beyond convexity significantly impedes a unified understanding of fundamental advantages and limitations of quantum mechanical properties, which general QRTs~\cite{Kuroiwa2020,PhysRevA.104.L020401,Chitambar2018,Horodecki2013b, Liu2017, Regula2017,Bromley2018,Anshu2018,Uola2019,Vijayan2019,Liu_ZW2019a,Takagi2019b,Takagi2019a,gonda2019monotones,Fang2020,Regula2020,Sparaciari2020,Kuroiwa2020,Brunner2021,Regula2021,Lami2021,Regula2021_fundamentallimitation,Regula2021one-shot,Regula2022,Fang2022,Regula2022tightconstraints,Lami2023,berta2023gap} aim to achieve.

\begin{figure}[t]
    \includegraphics[width = \columnwidth]{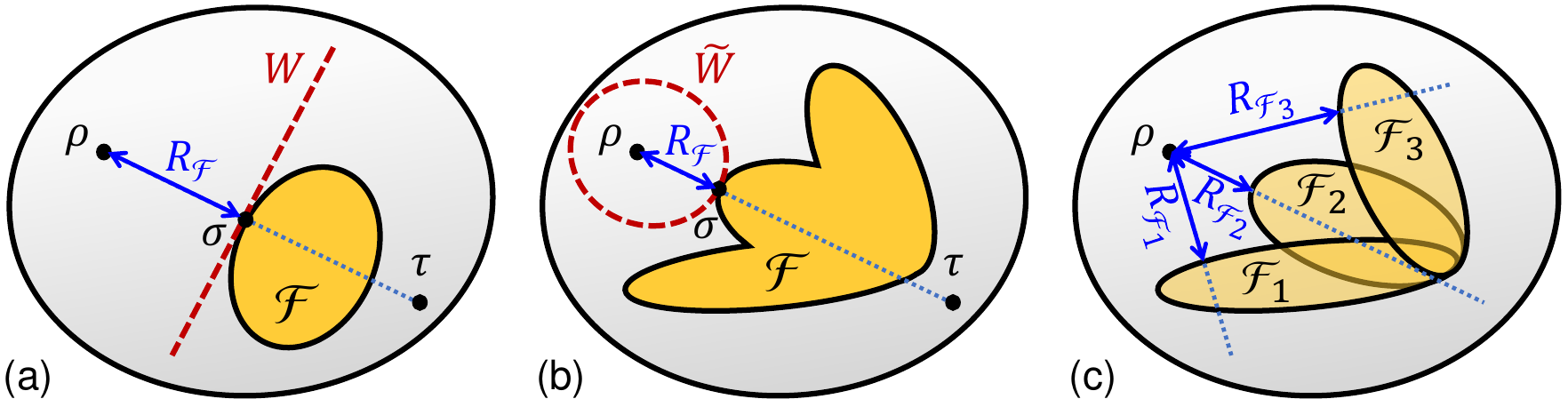}
   \caption{\rt{Generalized robustness $R_{\mathcal{F}}$ of a state $\rho$ in (a) convex and (b)--(c) non-convex QRTs [Def.~\ref{eq:robustness}]. The first two panels illustrate the link between $R_{\mathcal{F}}$ and witness operators, respectively for (a) the convex case~\cite{Takagi2019b} and (b) the non-convex case~[Thm.~\ref{thm:high_order_witness_qudit}]. Panel (c) depicts the evaluation of $R_{\mathcal{F}}$ as minimization over convex subsets $\mathcal{F}_k$ forming a non-convex set $\mathcal{F}$ of free states [Eq.~(\ref{eq:robustness_ink})]. Our constructions in (b) and (c) reveal qualitative [Thm.~\ref{thm:eqd}] and quantitative advantages [Thm.~\ref{thm:advantage worst case}] of all non-free states of general QRTs in channel discrimination tasks.}}
    \label{fig:robustness_prl}
\end{figure}

\rt{In this Letter, we bridge this gap by providing universal operational characterizations for general QRTs without convexity restriction, based on the generalized robustness (see Fig.~\ref{fig:robustness_prl}).}

On the one hand, we consider a concept of \textit{multi-copy resource witness}, extending ideas previously investigated in entanglement theory~\cite{Horodecki2003, Horodecki2009}. 
Resource witnesses are instrumental in detecting the usefulness of entanglement~\cite{Piani2009} and any convex resource~\cite{Takagi2019b}.
We provide a characterization of the generalized robustness in arbitrary dimensions and exploit it to give general constructions of multi-copy witnesses. This allows us to prove that {\it all} resource states in general QRTs (regardless of the topology of free set) produce an operational advantage in some multi-input channel discrimination task. 

On the other hand, we investigate a  scenario motivated by QRTs with possibly multiple constraints, that is, a case where the set of free states is composed in general of several different convex subsets.  We find that the generalized robustness quantifies the \textit{worst-case} advantage for channel discrimination with respect to each subset and thus captures the versatility of resource states in the presence of competing constraints. 

Our results show that, without assuming convexity, every quantum state identified as not free in a general QRT has the potential to lead to an advantage in discrimination problems, and the generalized robustness is established as a universal indicator of such operational advantage.
This answers the final question raised in the opening paragraph of this Letter.

\rt{In the following, we focus on exposing the main ideas and results concerning QRTs defined for quantum states. A companion paper~\cite{PRA} contains detailed proofs of the main results of this manuscript, more examples, and extensions to dynamical QRTs including quantum channels and instruments
\cite{Gour2019a,Liu_ZW2019b,Liu_YC2020,Li2018,Gour_Wilde2018,Takagi2020,gour2020dynamical,yuan2020oneshot,Regula2021_fundamentallimitation,Regula2021one-shot,Takagi2019a,Fang2022,Heinosaari2015_NoiseRobustness,Haapasalo2015_RobustnessIncompatibility,Guerini2017_MeasurmentSimulability,Skrzypczyk2019_RobustnessMeasurement,Skrzypczyk2019_IncompatibleMeasurements,Oszmaniec2019operational,Guff2021_ResourceMeasurement}, as well as other  tasks based on channel exclusion and weight-based resource measures \cite{Lewenstein1998,Skrzypczyk2014,Pusey2015,Cavalcanti2016,Kaifeng2018,Ducuara2020,Uola2020}.}

\rt{\textbf{\textit{Generalized robustness in general QRTs}}.---}
We consider general QRTs on a $d$-dimensional Hilbert space $\mathcal{H}$ (for finite $d$). 
The set $\mathcal{F}(\mathcal{H})$ of free states is defined as some closed subset of the set $\mathcal{D}(\mathcal{H})$ of all states of the system $\mathcal{H}$. 
The {\it generalized robustness} resource quantifier is defined as follows. 
\begin{definition}
    Let $\rho \in \mathcal{D}(\mathcal{H})$ be a quantum state. The generalized robustness $R_{\mathcal{F}(\mathcal{H})}(\rho)$ of $\rho$ with respect to the set $\mathcal{F}(\mathcal{H})$ of free states is defined by (see Fig.~\ref{fig:robustness_prl})
    \begin{equation}
        \label{eq:robustness}
        R_{\mathcal{F}(\mathcal{H})}(\rho) \coloneqq  \min_{\tau \in \mathcal{D}(\mathcal{H})}\left\{s\geq 0 : \dfrac{\rho + s \tau}{1 + s} \eqqcolon \sigma \in \mathcal{F}(\mathcal{H}) \right\}. 
    \end{equation}
\end{definition}
This captures how much the quantum state $\rho$ can tolerate mixing with some other state $\tau$ until all of its resource content is washed out. Note that we do not assume convexity of  $\mathcal{F}(\mathcal{H})$ while, conventionally, the notion of the generalized robustness has been studied for convex resource theories \cite{Steiner2003,Harrow2003,Napoli2016,Takagi2019b}.

\rt{A key feature underlying the operational interpretation of generalized robustness in convex QRTs \cite{Piani2009,Napoli2016,Takagi2019b} is the fact that such a measure can be recast as the expectation value of an optimal {\it linear} resource witness, represented by a (red dashed) {\it straight} line in Fig.~\ref{fig:robustness_prl}(a). Our first main result is to show that such a connection can still be established in non-convex QRTs, but it requires the use of a multi-copy witness, that is, a  {\it nonlinear} operator separating resource states from the generally non-convex set of free states, illustrated by the (red dashed) {\it curved} line in  Fig.~\ref{fig:robustness_prl}(b).} 

\textbf{\textit{Multi-copy resource witness based on generalized robustness}}.---
\rt{Here we construct a family of $m$-copy operators for each $m = 2,3,\ldots,d$ and then show that, for an arbitrary free state $\sigma$, there is at least one $m$ such that the corresponding $m$-copy witness can discern a given resource state $\rho$ from $\sigma$. Our construction relies on the generalized Bloch representation of $d$-dimensional quantum states~\cite{KIMURA2003339,Bertlmann_2008}.}  Given a quantum state $\eta \in \mathcal{D}(\mathcal{H})$, we let $\vec{x}\coloneqq(x_1,x_2,\ldots,x_{d^2-1})$ denote a generalized Bloch vector given by 
$\eta = \frac{1}{d}\left(I + \sqrt{\tfrac{d(d-1)}{2}} \sum_{j=1}^{d^2-1} x_j\lambda_j \right)$, 
where $I$ is the $d\times d$ identity matrix, and $\{\lambda_j\}_j$ denote the $d\times d$ generalized Gell-Mann matrices with normalization $\tr[\lambda_i^\dagger \lambda_j] = 2\delta_{ij}$~\cite{KIMURA2003339,Bertlmann_2008}. 
Take a resource state $\rho \in \mathcal{D}(\mathcal{H})\backslash \mathcal{F}(\mathcal{H})$ with generalized Bloch vector $\vec{r}\coloneqq(r_j)$. 
According to the characterization given in Ref.~\cite{Byrd2003_characterization}, for a fixed $s > 0$, any state $\eta$ with generalized Bloch vector $\vec x$
that can be written as $\eta = \frac{\rho + s'\tau}{1 + s'}$
using some $0 < s' \leq s$ and $\tau \in \mathcal{D}(\mathcal{H})$ must satisfy
    \begin{equation*}
        S_{m,\rho,s}(\eta) \coloneqq S_m\left(\frac{1+s}{s}\eta - \frac{1}{s}\rho \right) \geq 0
    \end{equation*}
for all $m = 1,2,\ldots,d$, where $S_m$ is defined recursively as
    $S_m(A) \coloneqq \frac{1}{m}\sum_{l=1}^{m} \left((-1)^{l-1}\tr\left[A^l\right]S_{m-l}(A)\right)$ 
for an operator $A$, with $m\geq 1$ and $S_0(A) \coloneqq 1$. 
Note that $S_1(A) = \tr[A]$, hence $S_{1,\rho,s}(\eta) =  1 \geq 0$ is trivially satisfied for any state $\eta \in \mathcal{D}(\mathcal{H})$. 

{By definition, $S_{m,\rho,s}$ is a degree-$m$ real-valued polynomial with respect to $\vec{x}$ for all $m = 1,2,\ldots,d$ and for any $s>0$. 
Hence, for an arbitrary choice of $m$ and $s$, there exists a Hermitian operator $W_m(\rho,s)$ on $\mathcal{H}^{\otimes m}$ such that $\tr\left[W_m(\rho,s)\eta^{\otimes m}\right] = S_{m,\rho,s}(\eta)$ 
for all $\eta \in \mathcal{D}(\mathcal{H})$; see~\cite{PRA} for an explicit construction. }
With this characterization, we show the following:  
\begin{enumerate}
    \item  $S_{m,\rho,s}(\rho) = S_m(\rho) \geq 0$ for all $m = 2,3,\ldots,d$. 
    \item If $s < R_{\mathcal{F}(\mathcal{H})}(\rho)$, then for any free state $\sigma \in \mathcal{F}(\mathcal{H})$, there exists $2 \leq m \leq d$ such that $S_{m,\rho,s}(\sigma) < 0$. 
    \item If $s \geq R_{\mathcal{F}(\mathcal{H})}(\rho)$, there exists a free state $\sigma \in \mathcal{F}(\mathcal{H})$ such that $S_{m,\rho,s}(\sigma) \geq 0$ for all $m = 2,3,\ldots, d$.
\end{enumerate} 

Thus, the family $\mathcal{W}_{\rho,s} \coloneqq (W_m(\rho,s):m=2,3,\ldots,d)$ of Hermitian operators satisfies the two conditions: {$\min_{m = 2,3,\ldots,d}\tr\left[W_{m}(\rho,s)\rho^{\otimes m}\right] \geq 0$, and $\min_{m = 2,3,\ldots,d} \tr\left[W_{m}(\rho,s)\sigma^{\otimes m}\right] < 0, \,\,\forall \sigma \in \mathcal{F}(\mathcal{H})$.} {By defining} $\widetilde{W}_m(\rho,s) \coloneqq -C\left(W(\rho,s) + \Delta_{m}(\rho,s) I^{\otimes m}\right)$ with appropriate {scalar} $\Delta_{m}(\rho,s) > 0$ and any normalization constant $C >0$, we have that the family $\widetilde{\mathcal{W}}_{\rho,s} \coloneqq(\widetilde{W}_m(\rho,s): m = 2,3,\ldots,d)$ defines a (multi-copy) resource witness for $s <  R_{\mathcal{F}(\mathcal{H})}(\rho)$.

\begin{theorem}
    \label{thm:high_order_witness_qudit}
    Let $\mathcal{H}$ be a $d$-dimensional Hilbert space. 
    Let $\rho \in \mathcal{D}(\mathcal{H})\backslash\mathcal{F}(\mathcal{H})$ be a resource state.  
    Then, we can construct a family $\widetilde{\mathcal{W}}_{\rho,s} \coloneqq(\widetilde{W}_m(\rho,s): m = 2,3,\ldots,d)$ of Hermitian operators such that 
\begin{equation}
    \label{eq:conventional_witness_resource}
    \begin{aligned}
    &\max_{m = 2,3,\ldots,d}\tr\left[\widetilde{W}_{m}(\rho,s)\rho^{\otimes m}\right] < 0, \\ 
    &\max_{m = 2,3,\ldots,d} \tr\left[\widetilde{W}_{m}(\rho,s)\sigma^{\otimes m}\right] \geq 0, \,\,\forall \sigma \in \mathcal{F}(\mathcal{H}),
\end{aligned}
\end{equation}
    if and only if $s < R_{\mathcal{F}(\mathcal{H})}(\rho)$.
    In particular, 
        \begin{equation*}\sup\left\{s>0: \widetilde{\mathcal{W}}_{\rho,s}\,\,\mathrm{is\,\,a\,\,witness} \right\} 
        = R_{\mathcal{F}(\mathcal{H})}(\rho).\end{equation*} 
\end{theorem}

\rt{This result establishes a fundamental connection between generalized robustness and multi-copy witnesses in general QRTs\@.}
Moreover, when $s$ grows closer to $R_{\mathcal{F}(\mathcal{H})}$, the family $\widetilde{\mathcal{W}}_{\rho,s}$ serves as a better witness, in the sense that it can detect more resource states. 
Indeed, if {$s'< s$}, then we see that \cite{PRA} 
    \begin{equation*}
        \begin{aligned}
        &\{\eta \in \mathcal{D}(\mathcal{H}): \tr[\widetilde{W}_m(\rho,s') \eta^{\otimes m}] < 0,\,\, \forall m=2,3,\ldots,d \} \\
        &\subsetneq \{\eta \in \mathcal{D}(\mathcal{H}): \tr[\widetilde{W}_m(\rho,s) \eta^{\otimes d}] < 0,\,\,\forall m=2,3,\ldots,d \}.
        \end{aligned}
    \end{equation*}

\textbf{\textit{Example: Single-Qubit QRTs}}.---
\rt{To illustrate our result,  let us consider the instance of single-qubit QRTs ($d = 2$).} 

In the single-qubit case, we can only focus on $S_{2}$. 
By definition, we have $S_2(A) = (\tr[A]^2 - \tr[A^2])/2$. 
Note that for a Hermitian operator $\eta$ with Bloch vector $\vec{x} \coloneqq (x_1,x_2,x_3)$, the condition $S_{2}(\eta) \geq 0$ can be written as
   $x_1^2 + x_2^2 + x_3^2 \leq 1$, 
whose region is equivalent to the Bloch ball. 

Fix a single-qubit state $\rho$ with Bloch vector $\vec{r}\coloneqq(r_1,r_2,r_3)$. 
For any quantum state $\eta$ with Bloch vector $\vec{x}$,  we have
\begin{equation}
        S_{2,\rho,s}(\eta) 
        = \frac{1}{4} - \frac{\|\vec{r}\|^2}{4s^2} + \dfrac{(1+s)(\vec{r}\cdot\vec{x})}{2s^2}
         - \dfrac{(1+s)^2\|\vec{x}\|^2}{4s^2}. 
\end{equation}

Now, we construct a $2$-copy witness $W(\rho,s)$ such that $\tr [W(\rho,s)\eta^{\otimes 2}] = S_{2,\rho,s}(\eta)$. 
One such witness is given as~\cite{PRA}
$W(\rho,s)
        \coloneqq \left(\frac{\|\vec{r}\|^2}{s^2} - 1\right)I\otimes I 
         + \frac{(1+s)}{s^2} \sum_{j=1}^3 \big[(1+s) (\sigma_j\otimes\sigma_j) - r_j 
        (I\otimes \sigma_j + \sigma_j\otimes I)\big]$,
where $(\sigma_1, \sigma_2, \sigma_3)$ are the Pauli matrices. 
\rt{The operator $W(\rho,s)$ defines a spherical boundary around the state $\rho$ within the Bloch sphere, with radius proportional to $s$. It amounts to a witness when $s$ is smaller than the generalized robustness of $\rho$. The larger $s$ is, the more effective $W(\rho,s)$ is as a witness. In the limit of $s\! \rightarrow \! R_{\mathcal{F}(\mathcal{H})}(\rho)$, the boundary is tangent to the free set ${\mathcal{F}(\mathcal{H})}$, denoting an optimal witness [see Fig.~\ref{fig:robustness_prl}(b)].}

\textbf{\textit{Operational advantage in multi-input channel discrimination}}.---
The multi-copy witness constructed in the previous section leads to an operational advantage of all resource states in general QRTs without convexity restriction for a variant of channel discrimination, which we call \textit{$m$-input channel discrimination}. 
In $m$-input channel discrimination, one aims to distinguish channels that act on $m$ independently and identically distributed (i.i.d.) copies $\rho^{\otimes m}$ of a given state $\rho$ in a black box setting. 
Let $\{p_i,\Lambda^{(m)}_i\}_i$ be an ensemble of channels, where channels $\Lambda^{(m)}_i$ on $\mathcal{D}(\mathcal{H}^{\otimes m})$ are randomly picked with prior probability $p_i$. 
Once a channel $\Lambda^{(m)}_i$ is sampled, it acts on  the $m$-copy register of the given probe state $\rho$; that is, at the output, we have $\Lambda^{(m)}_i(\rho^{\otimes m})$. 
Our goal is to figure out which specific channel from the channel ensemble acted on the input state. 
To identify the label $i$, we perform a quantum measurement $\{M_i\}_i$ to the output state. 
The success probability for this $m$-input channel discrimination task is given by 
\begin{equation}
    \label{eq:success_prob}
  \!\!\!  p_{\mathrm{succ}}\left(\left\{p_i,\Lambda^{(m)}_i\right\}_i, \{M_i\}_i, \rho^{\otimes m}\right) \coloneqq \sum_{i} p_i \tr\left[M_i~\Lambda^{(m)}_i(\rho^{\otimes m})\right].\!
\end{equation}
In this task, given a quantum state and a channel ensemble, 
we aim to maximize the success probability by choosing the best measurement strategy. 
This task generalizes a single-copy channel discrimination task in Ref.~\cite{Takagi2019b} to the multi-copy scenario.
To characterize the advantage of a resource state $\rho$ in this scenario, we consider the ratio between the best success probability when using the given state $\rho$ versus a free state.

Theorem~\ref{thm:high_order_witness_qudit} shows that for any given $d$-dimensional resource state $\rho$, we may construct a family $\widetilde{\mathcal{W}}_{\rho,s} = (\widetilde{W}_m(\rho,s):m=2,3,\ldots,d)$ such that for any free state $\sigma \in \mathcal{F}(\mathcal{H})$, at least one $\widetilde{W}_m(\rho,s)$ separates $\rho^{\otimes m}$ from $\sigma^{\otimes m}$. 
This separation implies that {\it all resource states} in a general QRT without convexity restriction enable an operational advantage, formalized as follows. 

\begin{theorem}\label{thm:eqd}
Let $\mathcal{H}$ be a $d$-dimensional Hilbert space. 
For any resource state $\rho \in \mathcal{D}(\mathcal{H})\backslash\mathcal{F}(\mathcal{H})$, there exists a family of channel ensembles $\left(\{p_i,\Lambda^{(m)}_i\}_i\right)_{m=2}^{d}$ such that 
\begin{equation*}
    \min_{\sigma \in \mathcal{F}(\mathcal{H})}\max_{m = 2,3,\ldots,d}\dfrac{\displaystyle \max_{\{M^{(m)}_i\}_i} p_{\mathrm{succ}}(\{p_i,\Lambda^{(m)}_i\}_i,\{M^{(m)}_i\}_i,\rho^{\otimes m})}{\displaystyle \max_{\{M^{(m)}_i\}_i} p_{\mathrm{succ}}(\{p_i,\Lambda^{(m)}_i\}_i,\{M^{(m)}_i\}_i,\sigma^{\otimes m})} > 1. 
\end{equation*}
\end{theorem}
\rt{This can be proven by considering an $m$-input channel discrimination with two channels for each $m = 2,3,\ldots,d$, 
based on the witness $\widetilde{\mathcal{W}}_{\rho,s}$. 
The proof strategy is inspired by those in Refs.~\cite{Piani2009} and \cite{Takagi2019b}, 
and a detailed proof is reported in~\cite{PRA}.}

\rt{
{\it Remark.} While our construction requires $m\leq d$ inputs to reveal a discrimination advantage based on the generalized robustness, experimentally friendlier witnesses can be crafted just for detecting resource states in non-convex QRTs. Consider the operator $W^{\prime}_2({\rho, \varepsilon}) \coloneqq V + (\tr[\rho^2]-\varepsilon)\, I \otimes I - 2\, \rho \otimes I$ that can be implemented on $2$ copies of $\rho$, where $\varepsilon>0$ and $V$ is the {\sc{swap}} operator, $V \ket{\psi} \otimes \ket{\phi} = \ket{\phi} \otimes \ket{\psi}$ \cite{Ekert2002}. We have $\tr[W^{\prime}_2({\rho, \varepsilon})\eta^{\otimes 2}] = \tr[(\rho-\eta)^2]-\varepsilon$ for any state $\eta$. This defines an $\varepsilon$-ball around $\rho$, yielding a universal resource witness that satisfies Eqs.~(\ref{eq:conventional_witness_resource}) with $m=2$ for any $d$, if  $\varepsilon$ is small enough so that $\tr[(\rho - \sigma)^2] \geq \varepsilon$ for all free states  $\sigma \in \mathcal{F}(\mathcal{H})$.
}

\textbf{\textit{Operational advantage quantified by generalized robustness}}.--- 
In the previous section, based on the multi-copy witness constructed from generalized robustness, we showed a {\it qualitative} advantage of all resource states for channel discrimination in the framework of non-convex QRTs. 
\rt{The next question is whether we can provide a {\it quantitative} assessment of the operational advantage in general. Here we address this question by considering a slight variation of the problem.} 

Suppose that the set $\mathcal{F}(\mathcal{H})$ of free states of an arbitrary QRT can  be expressed without loss of generality   as a union  
    \begin{equation}\label{eq:deconvex}
        \mathcal{F}(\mathcal{H}) = \bigcup_{k} \mathcal{F}_k(\mathcal{H}), 
    \end{equation}
where the subsets $\mathcal{F}_k(\mathcal{H})$ are closed and {\it convex} for all $k$ [Fig.~\ref{fig:robustness_prl}(c)], and $k$ is either a discrete or continuous label. 
Such a decomposition is always possible (and generally not unique) and describes a QRT characterized by multiple constraints. 

For example, in the theory of quantum discord \cite{Ollivier2001,Henderson2001}, the set of free `classical-quantum' states is given by the union of the convex subsets of incoherent-quantum states with a fixed local basis \cite{ABC2016}. From a practical perspective, for applications such as 3D magnetic field sensing \cite{Baumgratz2016}, one needs states with coherence in a specific set of bases, e.g., $x$, $y$, and $z$ bases for a qubit. In this case, the set of free states is the union of the set of incoherent states with respect to the given bases, covering the three main axes of the Bloch sphere, while any other state is a resource. In another context, thermodynamical machines such as engines or refrigerators operating between different thermal baths are specified by a set of free states given by the union of the convex subsets corresponding to each equilibrium temperature \cite{Correa2014, Korzekwa2023}. 
Also in the context of multipartite entanglement,  states in the union of the sets of partition-separable states with respect to different bipartitions cannot have genuine multipartite entanglement~\cite{Yamasaki2022activationofgenuine,Palazuelos2022genuinemultipartite}.
For any state out of this non-convex set, genuine multipartite entanglement can be activated from many copies of the state, characterizing this non-convex set from a resource perspective~\cite{Yamasaki2022activationofgenuine,Palazuelos2022genuinemultipartite}.

Now, given a general QRT  described by multiple constraints (i.e., multiple free subsets), how do we characterize the usefulness of resource states? It is natural to introduce a {\em worst-case} scenario in which we consider the possible advantage that a given state can provide in a task when compared to each subset of free states, and then minimize such advantage over all such subsets. This approach will result in a figure of merit assessing how useful that particular state is {\it a priori} guaranteed to be, regardless of which particular constraints are coming into play in each run of the task. An instance of such analysis had been explored in metrological contexts --- that is, looking for the worst-case precision that a probe state enables in phase estimation with limited prior knowledge of the phase shift generator --- and led to seminal operational interpretations for discord-like quantifiers \cite{Girolami2013,Girolami2014,Farace2014}, which can now be regarded as special cases of our more general construction. 

Here,  we consider a channel discrimination task specified by a channel ensemble $\{p_i,\Lambda_i\}_i$ acting on a single copy of an input state, corresponding to $m=1$ in the $m$-input protocol described earlier. We  focus on the achievable advantage of a resource state $\rho$ with respect to states in each convex free subset $\mathcal{F}_k(\mathcal{H})$, maximized over the measurement  $\{M_i\}_i$. In formula, 
\begin{equation}
    \label{eq:k-advantage}
\max_{\{p_i,\Lambda_i\}_i,\{M_i\}_i} \dfrac{p_{\mathrm{\mathrm{succ}}}(\rho,\{p_i,\Lambda_i\}_i,\{M_i\}_i)}{\displaystyle \max_{\sigma_k \in \mathcal{F}_k(\mathcal{H})} p_{\mathrm{\mathrm{succ}}}(\sigma_k,\{p_i,\Lambda_i\}_i,\{M_i\}_i)}, 
\end{equation}
where $p_{\mathrm{\mathrm{succ}}}$ denotes the success probability, which is given in Eq.~\eqref{eq:success_prob}. 
We then consider the worst-case advantage of $\rho$ defined as the infimum of Eq.~\eqref{eq:k-advantage} over all free subsets $\mathcal{F}_k(\mathcal{H})$.  

Remarkably, we find that such a worst-case advantage is exactly quantified by the generalized robustness of $\rho$, Eq.~(\ref{eq:robustness}).
\begin{theorem}\label{thm:advantage worst case}
    Let $\mathcal{H}$ be a $d$-dimensional Hilbert space. 
    Then, for any resource state $\rho \in \mathcal{F}(\mathcal{H})\backslash \mathcal{D}(\mathcal{H})$, 
    \begin{equation*}
        \begin{aligned}
            \inf_{k} \max_{\{p_i,\Lambda_i\}_i,\{M_i\}_i} \dfrac{p_{\mathrm{\mathrm{succ}}}(\rho,\{p_i,\Lambda_i\}_i,\{M_i\}_i)}{\displaystyle \max_{\sigma_k \in \mathcal{F}_k(\mathcal{H})} p_{\mathrm{\mathrm{succ}}}(\sigma_k,\{p_i,\Lambda_i\}_i,\{M_i\}_i)} 
            = 1 + R_{\mathcal{F}(\mathcal{H})}(\rho). 
        \end{aligned}
    \end{equation*}
\end{theorem} 
This result illustrates that the generalized robustness in a general QRT precisely quantifies the usefulness of resource states in worst-case channel discrimination tasks. Crucially, this result does not depend on the specifics of the decomposition of the set of free states into convex subsets, but it holds for any representation of the form (\ref{eq:deconvex}), therefore standing as a universal staple of QRTs. 
We remark that when the set $\mathcal{F}(\mathcal{H})$ of free states is convex, the infimum over $k$ can be omitted, and Theorem~\ref{thm:advantage worst case} recovers the known result for convex QRTs~\cite{Takagi2019b}.

\rt{We now outline the proof, which consists of two main ingredients: (i) the results of~\cite{Takagi2019b} for convex QRTs, and (ii) the generalized robustness in the worst-case scenario. 
Let us fix an index $k$, and consider the corresponding convex free subset $\mathcal{F}_{k}(\mathcal{H})$. 
We can apply the result for convex QRTs~\cite{Takagi2019b} to $\mathcal{F}_{k}(\mathcal{H})$; 
for any state $\rho \in \mathcal{D}(\mathcal{H})$, we have the characterization 
\begin{equation*}
            \max_{\{p_i,\Lambda_i\}_i,\{M_i\}_i} \dfrac{p_{\mathrm{\mathrm{succ}}}(\rho,\{p_i,\Lambda_i\}_i,\{M_i\}_i)}{\displaystyle\max_{\sigma_k \in \mathcal{F}_k(\mathcal{H})} p_{\mathrm{\mathrm{succ}}}(\sigma_k,\{p_i,\Lambda_i\}_i,\{M_i\}_i)}  
            = 1 + R_{\mathcal{F}_k(\mathcal{H})}(\rho). 
    \end{equation*}
Now, we consider the worst-case advantage; mathematically, we take the infimum over the index $k$ on both sides. 
It only remains to evaluate $\inf_k R_{\mathcal{F}}(\rho)$. 
Observing that an optimal free state $\sigma \in \mathcal{F}(\mathcal{H})$ should belong to one of the convex subsets $\mathcal{F}_{k}(\mathcal{H})$, we have the following identity, concluding the proof, 
\begin{equation}\label{eq:robustness_ink}
    \inf_k R_{\mathcal{F}_k(\mathcal{H})}(\rho) = R_{\mathcal{F}(\mathcal{H})}(\rho).
\end{equation}
}

\textbf{\textit{Conclusion}}.---
In this Letter, we tackled two outstanding problems: (i) Can every quantum state provide an advantage in informational tasks without the need to introduce additional constraints? and (ii) if yes, can such an advantage be quantified in general terms? We answered both questions in the affirmative by providing two different yet related characterizations of the generalized robustness in generally non-convex QRTs. 
First, we introduced a multi-copy resource witness in $d$-dimensional QRTs without convexity assumption, which guarantees that there exists a family of multi-input channel discrimination tasks in which all resource states show an operational advantage. 
Second, we showed that the generalized robustness exactly quantifies the operational advantage of a resource state in a single-copy channel discrimination task when a worst-case scenario is considered. In this setting, the non-convex set of free states can be expressed as the union of multiple convex subsets, and the generalized robustness amounts to the minimum of all possible advantages that a resource state can achieve  with respect to each free subset. 
{Taken together, our results enhance the domain of applications of QRTs and establish a fundamental stepping stone in the ongoing translation of quantum science into quantum technologies, by establishing that meaningful {\it quantumness} can be identified without any need for convexity requirements, and that such quantumness can be directly exploited for practical applications.}

\begin{acknowledgments}
We are grateful to Bartosz Regula and Marco Piani for discussions and advice on this project. 
We also acknowledge Pauli T. J. Jokinen for pointing out an error in the earlier version of this paper. 
K.K. was supported by a Mike and Ophelia Lazaridis Fellowship, a Funai Overseas Scholarship, and a Perimeter Residency Doctoral Award. 
R.T. acknowledges the support of JSPS KAKENHI Grant Number JP23K19028, and JST, CREST Grant Number JPMJCR23I3, Japan. 
G.A. acknowledges support by the UK Research and Innovation (UKRI) under BBSRC Grant No.~BB/X004317/1 and EPSRC Grant No.~EP/X010929/1.
H.Y. acknowledges JST PRESTO Grant Number JPMJPR201A, JPMJPR23FC, and MEXT Quantum Leap Flagship Program (MEXT QLEAP) JPMXS0118069605, JPMXS0120351339\@. A part of this work was carried out at the workshop ``Quantum resources: from mathematical foundations to operational characterisation'' held in Singapore in December 2022.
\end{acknowledgments}

\bibliographystyle{apsrmp4-2}
\bibliography{robustness}

\begin{thebibliography}{92}%
\makeatletter
\providecommand \@ifxundefined [1]{%
 \@ifx{#1\undefined}
}%
\providecommand \@ifnum [1]{%
 \ifnum #1\expandafter \@firstoftwo
 \else \expandafter \@secondoftwo
 \fi
}%
\providecommand \@ifx [1]{%
 \ifx #1\expandafter \@firstoftwo
 \else \expandafter \@secondoftwo
 \fi
}%
\providecommand \natexlab [1]{#1}%
\providecommand \emph  [1]{``#1''}%
\providecommand \bibnamefont  [1]{#1}%
\providecommand \bibfnamefont [1]{#1}%
\providecommand \citenamefont [1]{#1}%
\providecommand \href@noop [0]{\@secondoftwo}%
\providecommand \href [0]{\begingroup \@sanitize@url \@href}%
\providecommand \@href[1]{\@@startlink{#1}\@@href}%
\providecommand \@@href[1]{\endgroup#1\@@endlink}%
\providecommand \@sanitize@url [0]{\catcode `\\12\catcode `\$12\catcode
  `\&12\catcode `\#12\catcode `\^12\catcode `\_12\catcode `\%12\relax}%
\providecommand \@@startlink[1]{}%
\providecommand \@@endlink[0]{}%
\providecommand \url  [0]{\begingroup\@sanitize@url \@url }%
\providecommand \@url [1]{\endgroup\@href {#1}{\urlprefix }}%
\providecommand \urlprefix  [0]{URL }%
\providecommand \Eprint [0]{\href }%
\providecommand \doibase [0]{http://dx.doi.org/}%
\providecommand \selectlanguage [0]{\@gobble}%
\providecommand \bibinfo  [0]{\@secondoftwo}%
\providecommand \bibfield  [0]{\@secondoftwo}%
\providecommand \translation [1]{[#1]}%
\providecommand \BibitemOpen [0]{}%
\providecommand \bibitemStop [0]{}%
\providecommand \bibitemNoStop [0]{.\EOS\space}%
\providecommand \EOS [0]{\spacefactor3000\relax}%
\providecommand \BibitemShut  [1]{\csname bibitem#1\endcsname}%
\let\auto@bib@innerbib\@empty
%</preamble>
\bibitem [{\citenamefont {Schirber}(2022)}]{Nobel2022}%
  \BibitemOpen
  \bibfield  {author} {\bibinfo {author} {\bibfnamefont {M.}~\bibnamefont
  {Schirber}},\ }\bibfield  {title} {\emph {\bibinfo {title} {Nobel prize:
  Quantum entanglement unveiled},}\ }\href
  {http://dx.doi.org/10.1103/Physics.15.153} {\bibfield  {journal} {\bibinfo
  {journal} {Physics}\ }\textbf {\bibinfo {volume} {15}},\ \bibinfo {pages}
  {153} (\bibinfo {year} {2022})}\BibitemShut {NoStop}%
\bibitem [{\citenamefont {Dowling}\ and\ \citenamefont
  {Milburn}(2003)}]{Dowling2003}%
  \BibitemOpen
  \bibfield  {author} {\bibinfo {author} {\bibfnamefont {J.~P.}\ \bibnamefont
  {Dowling}}\ and\ \bibinfo {author} {\bibfnamefont {G.~J.}\ \bibnamefont
  {Milburn}},\ }\bibfield  {title} {\emph {\bibinfo {title} {Quantum
  technology: the second quantum revolution},}\ }\href
  {http://dx.doi.org/10.1098/rsta.2003.1227} {\bibfield  {journal} {\bibinfo
  {journal} {Philos. Trans. R. Soc. Lond. Ser. A: Math. Phys. Eng. Sci.}\
  }\textbf {\bibinfo {volume} {361}},\ \bibinfo {pages} {1655} (\bibinfo {year}
  {2003})}\BibitemShut {NoStop}%
\bibitem [{\citenamefont {Streltsov}\ \emph {et~al.}(2017)\citenamefont
  {Streltsov}, \citenamefont {Adesso},\ and\ \citenamefont
  {Plenio}}]{Streltsov2017}%
  \BibitemOpen
  \bibfield  {author} {\bibinfo {author} {\bibfnamefont {A.}~\bibnamefont
  {Streltsov}}, \bibinfo {author} {\bibfnamefont {G.}~\bibnamefont {Adesso}}, \
  and\ \bibinfo {author} {\bibfnamefont {M.~B.}\ \bibnamefont {Plenio}},\
  }\bibfield  {title} {\emph {\bibinfo {title} {Colloquium: Quantum coherence
  as a resource},}\ }\href {http://dx.doi.org/10.1103/RevModPhys.89.041003}
  {\bibfield  {journal} {\bibinfo  {journal} {Rev. Mod. Phys.}\ }\textbf
  {\bibinfo {volume} {89}},\ \bibinfo {pages} {041003} (\bibinfo {year}
  {2017})}\BibitemShut {NoStop}%
\bibitem [{\citenamefont {Chitambar}\ and\ \citenamefont
  {Gour}(2019)}]{Chitambar2018}%
  \BibitemOpen
  \bibfield  {author} {\bibinfo {author} {\bibfnamefont {E.}~\bibnamefont
  {Chitambar}}\ and\ \bibinfo {author} {\bibfnamefont {G.}~\bibnamefont
  {Gour}},\ }\bibfield  {title} {\emph {\bibinfo {title} {Quantum resource
  theories},}\ }\href {http://dx.doi.org/10.1103/RevModPhys.91.025001}
  {\bibfield  {journal} {\bibinfo  {journal} {Rev. Mod. Phys.}\ }\textbf
  {\bibinfo {volume} {91}},\ \bibinfo {pages} {025001} (\bibinfo {year}
  {2019})}\BibitemShut {NoStop}%
\bibitem [{\citenamefont {Kuroiwa}\ and\ \citenamefont
  {Yamasaki}(2020)}]{Kuroiwa2020}%
  \BibitemOpen
  \bibfield  {author} {\bibinfo {author} {\bibfnamefont {K.}~\bibnamefont
  {Kuroiwa}}\ and\ \bibinfo {author} {\bibfnamefont {H.}~\bibnamefont
  {Yamasaki}},\ }\bibfield  {title} {\emph {\bibinfo {title} {General {Q}uantum
  {R}esource {T}heories: {D}istillation, {F}ormation and {C}onsistent
  {R}esource {M}easures},}\ }\href
  {http://dx.doi.org/10.22331/q-2020-11-01-355} {\bibfield  {journal} {\bibinfo
   {journal} {{Quantum}}\ }\textbf {\bibinfo {volume} {4}},\ \bibinfo {pages}
  {355} (\bibinfo {year} {2020})}\BibitemShut {NoStop}%
\bibitem [{\citenamefont {Vidal}\ and\ \citenamefont
  {Tarrach}(1999)}]{Vidal1999}%
  \BibitemOpen
  \bibfield  {author} {\bibinfo {author} {\bibfnamefont {G.}~\bibnamefont
  {Vidal}}\ and\ \bibinfo {author} {\bibfnamefont {R.}~\bibnamefont
  {Tarrach}},\ }\bibfield  {title} {\emph {\bibinfo {title} {Robustness of
  entanglement},}\ }\href {http://dx.doi.org/10.1103/PhysRevA.59.141}
  {\bibfield  {journal} {\bibinfo  {journal} {Phys. Rev. A}\ }\textbf {\bibinfo
  {volume} {59}},\ \bibinfo {pages} {141} (\bibinfo {year} {1999})}\BibitemShut
  {NoStop}%
\bibitem [{\citenamefont {Steiner}(2003)}]{Steiner2003}%
  \BibitemOpen
  \bibfield  {author} {\bibinfo {author} {\bibfnamefont {M.}~\bibnamefont
  {Steiner}},\ }\bibfield  {title} {\emph {\bibinfo {title} {Generalized
  robustness of entanglement},}\ }\href
  {http://dx.doi.org/10.1103/PhysRevA.67.054305} {\bibfield  {journal}
  {\bibinfo  {journal} {Phys. Rev. A}\ }\textbf {\bibinfo {volume} {67}},\
  \bibinfo {pages} {054305} (\bibinfo {year} {2003})}\BibitemShut {NoStop}%
\bibitem [{\citenamefont {Harrow}\ and\ \citenamefont
  {Nielsen}(2003)}]{Harrow2003}%
  \BibitemOpen
  \bibfield  {author} {\bibinfo {author} {\bibfnamefont {A.~W.}\ \bibnamefont
  {Harrow}}\ and\ \bibinfo {author} {\bibfnamefont {M.~A.}\ \bibnamefont
  {Nielsen}},\ }\bibfield  {title} {\emph {\bibinfo {title} {Robustness of
  quantum gates in the presence of noise},}\ }\href
  {http://dx.doi.org/10.1103/PhysRevA.68.012308} {\bibfield  {journal}
  {\bibinfo  {journal} {Phys. Rev. A}\ }\textbf {\bibinfo {volume} {68}},\
  \bibinfo {pages} {012308} (\bibinfo {year} {2003})}\BibitemShut {NoStop}%
\bibitem [{\citenamefont {Bae}\ \emph {et~al.}(2019)\citenamefont {Bae},
  \citenamefont {Chru\ifmmode \acute{s}\else
  \'{s}\fi{}ci\ifmmode~\acute{n}\else \'{n}\fi{}ski},\ and\ \citenamefont
  {Piani}}]{Bae2019}%
  \BibitemOpen
  \bibfield  {author} {\bibinfo {author} {\bibfnamefont {J.}~\bibnamefont
  {Bae}}, \bibinfo {author} {\bibfnamefont {D.}~\bibnamefont {Chru\ifmmode
  \acute{s}\else \'{s}\fi{}ci\ifmmode~\acute{n}\else \'{n}\fi{}ski}}, \ and\
  \bibinfo {author} {\bibfnamefont {M.}~\bibnamefont {Piani}},\ }\bibfield
  {title} {\emph {\bibinfo {title} {More entanglement implies higher
  performance in channel discrimination tasks},}\ }\href
  {http://dx.doi.org/10.1103/PhysRevLett.122.140404} {\bibfield  {journal}
  {\bibinfo  {journal} {Phys. Rev. Lett.}\ }\textbf {\bibinfo {volume} {122}},\
  \bibinfo {pages} {140404} (\bibinfo {year} {2019})}\BibitemShut {NoStop}%
\bibitem [{\citenamefont {Napoli}\ \emph {et~al.}(2016)\citenamefont {Napoli},
  \citenamefont {Bromley}, \citenamefont {Cianciaruso}, \citenamefont {Piani},
  \citenamefont {Johnston},\ and\ \citenamefont {Adesso}}]{Napoli2016}%
  \BibitemOpen
  \bibfield  {author} {\bibinfo {author} {\bibfnamefont {C.}~\bibnamefont
  {Napoli}}, \bibinfo {author} {\bibfnamefont {T.~R.}\ \bibnamefont {Bromley}},
  \bibinfo {author} {\bibfnamefont {M.}~\bibnamefont {Cianciaruso}}, \bibinfo
  {author} {\bibfnamefont {M.}~\bibnamefont {Piani}}, \bibinfo {author}
  {\bibfnamefont {N.}~\bibnamefont {Johnston}}, \ and\ \bibinfo {author}
  {\bibfnamefont {G.}~\bibnamefont {Adesso}},\ }\bibfield  {title} {\emph
  {\bibinfo {title} {Robustness of coherence: An operational and observable
  measure of quantum coherence},}\ }\href
  {http://dx.doi.org/10.1103/PhysRevLett.116.150502} {\bibfield  {journal}
  {\bibinfo  {journal} {Phys. Rev. Lett.}\ }\textbf {\bibinfo {volume} {116}},\
  \bibinfo {pages} {150502} (\bibinfo {year} {2016})}\BibitemShut {NoStop}%
\bibitem [{\citenamefont {Ringbauer}\ \emph {et~al.}(2018)\citenamefont
  {Ringbauer}, \citenamefont {Bromley}, \citenamefont {Cianciaruso},
  \citenamefont {Lami}, \citenamefont {Lau}, \citenamefont {Adesso},
  \citenamefont {White}, \citenamefont {Fedrizzi},\ and\ \citenamefont
  {Piani}}]{Ringbauer2018}%
  \BibitemOpen
  \bibfield  {author} {\bibinfo {author} {\bibfnamefont {M.}~\bibnamefont
  {Ringbauer}}, \bibinfo {author} {\bibfnamefont {T.~R.}\ \bibnamefont
  {Bromley}}, \bibinfo {author} {\bibfnamefont {M.}~\bibnamefont
  {Cianciaruso}}, \bibinfo {author} {\bibfnamefont {L.}~\bibnamefont {Lami}},
  \bibinfo {author} {\bibfnamefont {W.~Y.~S.}\ \bibnamefont {Lau}}, \bibinfo
  {author} {\bibfnamefont {G.}~\bibnamefont {Adesso}}, \bibinfo {author}
  {\bibfnamefont {A.~G.}\ \bibnamefont {White}}, \bibinfo {author}
  {\bibfnamefont {A.}~\bibnamefont {Fedrizzi}}, \ and\ \bibinfo {author}
  {\bibfnamefont {M.}~\bibnamefont {Piani}},\ }\bibfield  {title} {\emph
  {\bibinfo {title} {Certification and quantification of multilevel quantum
  coherence},}\ }\href {http://dx.doi.org/10.1103/PhysRevX.8.041007} {\bibfield
   {journal} {\bibinfo  {journal} {Phys. Rev. X}\ }\textbf {\bibinfo {volume}
  {8}},\ \bibinfo {pages} {041007} (\bibinfo {year} {2018})}\BibitemShut
  {NoStop}%
\bibitem [{\citenamefont {Piani}\ \emph {et~al.}(2016)\citenamefont {Piani},
  \citenamefont {Cianciaruso}, \citenamefont {Bromley}, \citenamefont {Napoli},
  \citenamefont {Johnston},\ and\ \citenamefont {Adesso}}]{Piani2016}%
  \BibitemOpen
  \bibfield  {author} {\bibinfo {author} {\bibfnamefont {M.}~\bibnamefont
  {Piani}}, \bibinfo {author} {\bibfnamefont {M.}~\bibnamefont {Cianciaruso}},
  \bibinfo {author} {\bibfnamefont {T.~R.}\ \bibnamefont {Bromley}}, \bibinfo
  {author} {\bibfnamefont {C.}~\bibnamefont {Napoli}}, \bibinfo {author}
  {\bibfnamefont {N.}~\bibnamefont {Johnston}}, \ and\ \bibinfo {author}
  {\bibfnamefont {G.}~\bibnamefont {Adesso}},\ }\bibfield  {title} {\emph
  {\bibinfo {title} {Robustness of asymmetry and coherence of quantum
  states},}\ }\href {http://dx.doi.org/10.1103/PhysRevA.93.042107} {\bibfield
  {journal} {\bibinfo  {journal} {Phys. Rev. A}\ }\textbf {\bibinfo {volume}
  {93}},\ \bibinfo {pages} {042107} (\bibinfo {year} {2016})}\BibitemShut
  {NoStop}%
\bibitem [{\citenamefont {Liu}\ \emph {et~al.}(2019)\citenamefont {Liu},
  \citenamefont {Bu},\ and\ \citenamefont {Takagi}}]{Liu_ZW2019a}%
  \BibitemOpen
  \bibfield  {author} {\bibinfo {author} {\bibfnamefont {Z.-W.}\ \bibnamefont
  {Liu}}, \bibinfo {author} {\bibfnamefont {K.}~\bibnamefont {Bu}}, \ and\
  \bibinfo {author} {\bibfnamefont {R.}~\bibnamefont {Takagi}},\ }\bibfield
  {title} {\emph {\bibinfo {title} {One-shot operational quantum resource
  theory},}\ }\href {http://dx.doi.org/10.1103/PhysRevLett.123.020401}
  {\bibfield  {journal} {\bibinfo  {journal} {Phys. Rev. Lett.}\ }\textbf
  {\bibinfo {volume} {123}},\ \bibinfo {pages} {020401} (\bibinfo {year}
  {2019})}\BibitemShut {NoStop}%
\bibitem [{\citenamefont {Seddon}\ \emph {et~al.}(2021)\citenamefont {Seddon},
  \citenamefont {Regula}, \citenamefont {Pashayan}, \citenamefont {Ouyang},\
  and\ \citenamefont {Campbell}}]{Seddon2021}%
  \BibitemOpen
  \bibfield  {author} {\bibinfo {author} {\bibfnamefont {J.~R.}\ \bibnamefont
  {Seddon}}, \bibinfo {author} {\bibfnamefont {B.}~\bibnamefont {Regula}},
  \bibinfo {author} {\bibfnamefont {H.}~\bibnamefont {Pashayan}}, \bibinfo
  {author} {\bibfnamefont {Y.}~\bibnamefont {Ouyang}}, \ and\ \bibinfo {author}
  {\bibfnamefont {E.~T.}\ \bibnamefont {Campbell}},\ }\bibfield  {title} {\emph
  {\bibinfo {title} {Quantifying quantum speedups: Improved classical
  simulation from tighter magic monotones},}\ }\href
  {http://dx.doi.org/10.1103/PRXQuantum.2.010345} {\bibfield  {journal}
  {\bibinfo  {journal} {PRX Quantum}\ }\textbf {\bibinfo {volume} {2}},\
  \bibinfo {pages} {010345} (\bibinfo {year} {2021})}\BibitemShut {NoStop}%
\bibitem [{\citenamefont {Piani}\ and\ \citenamefont
  {Watrous}(2015)}]{Piani2015}%
  \BibitemOpen
  \bibfield  {author} {\bibinfo {author} {\bibfnamefont {M.}~\bibnamefont
  {Piani}}\ and\ \bibinfo {author} {\bibfnamefont {J.}~\bibnamefont
  {Watrous}},\ }\bibfield  {title} {\emph {\bibinfo {title} {Necessary and
  sufficient quantum information characterization of einstein-podolsky-rosen
  steering},}\ }\href {http://dx.doi.org/10.1103/PhysRevLett.114.060404}
  {\bibfield  {journal} {\bibinfo  {journal} {Phys. Rev. Lett.}\ }\textbf
  {\bibinfo {volume} {114}},\ \bibinfo {pages} {060404} (\bibinfo {year}
  {2015})}\BibitemShut {NoStop}%
\bibitem [{\citenamefont {Takagi}\ \emph {et~al.}(2019)\citenamefont {Takagi},
  \citenamefont {Regula}, \citenamefont {Bu}, \citenamefont {Liu},\ and\
  \citenamefont {Adesso}}]{Takagi2019b}%
  \BibitemOpen
  \bibfield  {author} {\bibinfo {author} {\bibfnamefont {R.}~\bibnamefont
  {Takagi}}, \bibinfo {author} {\bibfnamefont {B.}~\bibnamefont {Regula}},
  \bibinfo {author} {\bibfnamefont {K.}~\bibnamefont {Bu}}, \bibinfo {author}
  {\bibfnamefont {Z.-W.}\ \bibnamefont {Liu}}, \ and\ \bibinfo {author}
  {\bibfnamefont {G.}~\bibnamefont {Adesso}},\ }\bibfield  {title} {\emph
  {\bibinfo {title} {Operational advantage of quantum resources in subchannel
  discrimination},}\ }\href {http://dx.doi.org/10.1103/PhysRevLett.122.140402}
  {\bibfield  {journal} {\bibinfo  {journal} {Phys. Rev. Lett.}\ }\textbf
  {\bibinfo {volume} {122}},\ \bibinfo {pages} {140402} (\bibinfo {year}
  {2019})}\BibitemShut {NoStop}%
\bibitem [{\citenamefont {Kitaev}(1997)}]{Kitaev_1997}%
  \BibitemOpen
  \bibfield  {author} {\bibinfo {author} {\bibfnamefont {A.~Y.}\ \bibnamefont
  {Kitaev}},\ }\bibfield  {title} {\emph {\bibinfo {title} {Quantum
  computations: algorithms and error correction},}\ }\href
  {http://dx.doi.org/10.1070/RM1997v052n06ABEH002155} {\bibfield  {journal}
  {\bibinfo  {journal} {Russ. Math. Surv.}\ }\textbf {\bibinfo {volume} {52}},\
  \bibinfo {pages} {1191} (\bibinfo {year} {1997})}\BibitemShut {NoStop}%
\bibitem [{\citenamefont {Childs}\ \emph {et~al.}(2000)\citenamefont {Childs},
  \citenamefont {Preskill},\ and\ \citenamefont {Renes}}]{Andrew2000}%
  \BibitemOpen
  \bibfield  {author} {\bibinfo {author} {\bibfnamefont {A.~M.}\ \bibnamefont
  {Childs}}, \bibinfo {author} {\bibfnamefont {J.}~\bibnamefont {Preskill}}, \
  and\ \bibinfo {author} {\bibfnamefont {J.}~\bibnamefont {Renes}},\ }\bibfield
   {title} {\emph {\bibinfo {title} {Quantum information and precision
  measurement},}\ }\href {http://dx.doi.org/10.1080/09500340008244034}
  {\bibfield  {journal} {\bibinfo  {journal} {J. Mod. Opt.}\ }\textbf {\bibinfo
  {volume} {47}},\ \bibinfo {pages} {155} (\bibinfo {year} {2000})}\BibitemShut
  {NoStop}%
\bibitem [{\citenamefont {Ac\'{\i}n}(2001)}]{Acin2001}%
  \BibitemOpen
  \bibfield  {author} {\bibinfo {author} {\bibfnamefont {A.}~\bibnamefont
  {Ac\'{\i}n}},\ }\bibfield  {title} {\emph {\bibinfo {title} {Statistical
  distinguishability between unitary operations},}\ }\href
  {http://dx.doi.org/10.1103/PhysRevLett.87.177901} {\bibfield  {journal}
  {\bibinfo  {journal} {Phys. Rev. Lett.}\ }\textbf {\bibinfo {volume} {87}},\
  \bibinfo {pages} {177901} (\bibinfo {year} {2001})}\BibitemShut {NoStop}%
\bibitem [{\citenamefont {Watrous}(2018)}]{watrous_2018}%
  \BibitemOpen
  \bibfield  {author} {\bibinfo {author} {\bibfnamefont {J.}~\bibnamefont
  {Watrous}},\ }\href {http://dx.doi.org/10.1017/9781316848142} {\emph
  {\bibinfo {title} {The Theory of Quantum Information}}}\ (\bibinfo
  {publisher} {Cambridge University Press},\ \bibinfo {year}
  {2018})\BibitemShut {NoStop}%
\bibitem [{\citenamefont {Takagi}\ and\ \citenamefont
  {Regula}(2019)}]{Takagi2019a}%
  \BibitemOpen
  \bibfield  {author} {\bibinfo {author} {\bibfnamefont {R.}~\bibnamefont
  {Takagi}}\ and\ \bibinfo {author} {\bibfnamefont {B.}~\bibnamefont
  {Regula}},\ }\bibfield  {title} {\emph {\bibinfo {title} {General resource
  theories in quantum mechanics and beyond: Operational characterization via
  discrimination tasks},}\ }\href {http://dx.doi.org/10.1103/PhysRevX.9.031053}
  {\bibfield  {journal} {\bibinfo  {journal} {Phys. Rev. X}\ }\textbf {\bibinfo
  {volume} {9}},\ \bibinfo {pages} {031053} (\bibinfo {year}
  {2019})}\BibitemShut {NoStop}%
\bibitem [{\citenamefont {Uola}\ \emph {et~al.}(2019)\citenamefont {Uola},
  \citenamefont {Kraft}, \citenamefont {Shang}, \citenamefont {Yu},\ and\
  \citenamefont {G\"uhne}}]{Uola2019}%
  \BibitemOpen
  \bibfield  {author} {\bibinfo {author} {\bibfnamefont {R.}~\bibnamefont
  {Uola}}, \bibinfo {author} {\bibfnamefont {T.}~\bibnamefont {Kraft}},
  \bibinfo {author} {\bibfnamefont {J.}~\bibnamefont {Shang}}, \bibinfo
  {author} {\bibfnamefont {X.-D.}\ \bibnamefont {Yu}}, \ and\ \bibinfo {author}
  {\bibfnamefont {O.}~\bibnamefont {G\"uhne}},\ }\bibfield  {title} {\emph
  {\bibinfo {title} {Quantifying quantum resources with conic programming},}\
  }\href {http://dx.doi.org/10.1103/PhysRevLett.122.130404} {\bibfield
  {journal} {\bibinfo  {journal} {Phys. Rev. Lett.}\ }\textbf {\bibinfo
  {volume} {122}},\ \bibinfo {pages} {130404} (\bibinfo {year}
  {2019})}\BibitemShut {NoStop}%
\bibitem [{\citenamefont {Regula}\ \emph {et~al.}(2021)\citenamefont {Regula},
  \citenamefont {Lami}, \citenamefont {Ferrari},\ and\ \citenamefont
  {Takagi}}]{Regula2021}%
  \BibitemOpen
  \bibfield  {author} {\bibinfo {author} {\bibfnamefont {B.}~\bibnamefont
  {Regula}}, \bibinfo {author} {\bibfnamefont {L.}~\bibnamefont {Lami}},
  \bibinfo {author} {\bibfnamefont {G.}~\bibnamefont {Ferrari}}, \ and\
  \bibinfo {author} {\bibfnamefont {R.}~\bibnamefont {Takagi}},\ }\bibfield
  {title} {\emph {\bibinfo {title} {Operational quantification of
  continuous-variable quantum resources},}\ }\href
  {http://dx.doi.org/10.1103/PhysRevLett.126.110403} {\bibfield  {journal}
  {\bibinfo  {journal} {Phys. Rev. Lett.}\ }\textbf {\bibinfo {volume} {126}},\
  \bibinfo {pages} {110403} (\bibinfo {year} {2021})}\BibitemShut {NoStop}%
\bibitem [{\citenamefont {Lami}\ \emph {et~al.}(2021)\citenamefont {Lami},
  \citenamefont {Regula}, \citenamefont {Takagi},\ and\ \citenamefont
  {Ferrari}}]{Lami2021}%
  \BibitemOpen
  \bibfield  {author} {\bibinfo {author} {\bibfnamefont {L.}~\bibnamefont
  {Lami}}, \bibinfo {author} {\bibfnamefont {B.}~\bibnamefont {Regula}},
  \bibinfo {author} {\bibfnamefont {R.}~\bibnamefont {Takagi}}, \ and\ \bibinfo
  {author} {\bibfnamefont {G.}~\bibnamefont {Ferrari}},\ }\bibfield  {title}
  {\emph {\bibinfo {title} {Framework for resource quantification in
  infinite-dimensional general probabilistic theories},}\ }\href
  {http://dx.doi.org/10.1103/PhysRevA.103.032424} {\bibfield  {journal}
  {\bibinfo  {journal} {Phys. Rev. A}\ }\textbf {\bibinfo {volume} {103}},\
  \bibinfo {pages} {032424} (\bibinfo {year} {2021})}\BibitemShut {NoStop}%
\bibitem [{\citenamefont {Brand\~ao}(2005)}]{Brandao2005}%
  \BibitemOpen
  \bibfield  {author} {\bibinfo {author} {\bibfnamefont {F.~G. S.~L.}\
  \bibnamefont {Brand\~ao}},\ }\bibfield  {title} {\emph {\bibinfo {title}
  {Quantifying entanglement with witness operators},}\ }\href
  {http://dx.doi.org/10.1103/PhysRevA.72.022310} {\bibfield  {journal}
  {\bibinfo  {journal} {Phys. Rev. A}\ }\textbf {\bibinfo {volume} {72}},\
  \bibinfo {pages} {022310} (\bibinfo {year} {2005})}\BibitemShut {NoStop}%
\bibitem [{\citenamefont {Eisert}\ \emph {et~al.}(2007)\citenamefont {Eisert},
  \citenamefont {Brandão},\ and\ \citenamefont {Audenaert}}]{Eisert_2007}%
  \BibitemOpen
  \bibfield  {author} {\bibinfo {author} {\bibfnamefont {J.}~\bibnamefont
  {Eisert}}, \bibinfo {author} {\bibfnamefont {F.~G. S.~L.}\ \bibnamefont
  {Brandão}}, \ and\ \bibinfo {author} {\bibfnamefont {K.~M.~R.}\ \bibnamefont
  {Audenaert}},\ }\bibfield  {title} {\emph {\bibinfo {title} {Quantitative
  entanglement witnesses},}\ }\href
  {http://dx.doi.org/10.1088/1367-2630/9/3/046} {\bibfield  {journal} {\bibinfo
   {journal} {New J. Phys.}\ }\textbf {\bibinfo {volume} {9}},\ \bibinfo
  {pages} {46} (\bibinfo {year} {2007})}\BibitemShut {NoStop}%
\bibitem [{\citenamefont {Walschaers}(2021)}]{Mattia2021}%
  \BibitemOpen
  \bibfield  {author} {\bibinfo {author} {\bibfnamefont {M.}~\bibnamefont
  {Walschaers}},\ }\bibfield  {title} {\emph {\bibinfo {title} {Non-gaussian
  quantum states and where to find them},}\ }\href
  {http://dx.doi.org/10.1103/PRXQuantum.2.030204} {\bibfield  {journal}
  {\bibinfo  {journal} {PRX Quantum}\ }\textbf {\bibinfo {volume} {2}},\
  \bibinfo {pages} {030204} (\bibinfo {year} {2021})}\BibitemShut {NoStop}%
\bibitem [{\citenamefont {Weedbrook}\ \emph {et~al.}(2012)\citenamefont
  {Weedbrook}, \citenamefont {Pirandola}, \citenamefont {Garc\'{\i}a-Patr\'on},
  \citenamefont {Cerf}, \citenamefont {Ralph}, \citenamefont {Shapiro},\ and\
  \citenamefont {Lloyd}}]{Weedbrook2012}%
  \BibitemOpen
  \bibfield  {author} {\bibinfo {author} {\bibfnamefont {C.}~\bibnamefont
  {Weedbrook}}, \bibinfo {author} {\bibfnamefont {S.}~\bibnamefont
  {Pirandola}}, \bibinfo {author} {\bibfnamefont {R.}~\bibnamefont
  {Garc\'{\i}a-Patr\'on}}, \bibinfo {author} {\bibfnamefont {N.~J.}\
  \bibnamefont {Cerf}}, \bibinfo {author} {\bibfnamefont {T.~C.}\ \bibnamefont
  {Ralph}}, \bibinfo {author} {\bibfnamefont {J.~H.}\ \bibnamefont {Shapiro}},
  \ and\ \bibinfo {author} {\bibfnamefont {S.}~\bibnamefont {Lloyd}},\
  }\bibfield  {title} {\emph {\bibinfo {title} {{Gaussian} quantum
  information},}\ }\href {http://dx.doi.org/10.1103/RevModPhys.84.621}
  {\bibfield  {journal} {\bibinfo  {journal} {Rev. Mod. Phys.}\ }\textbf
  {\bibinfo {volume} {84}},\ \bibinfo {pages} {621} (\bibinfo {year}
  {2012})}\BibitemShut {NoStop}%
\bibitem [{\citenamefont {Adesso}\ \emph {et~al.}(2014)\citenamefont {Adesso},
  \citenamefont {Ragy},\ and\ \citenamefont {Lee}}]{Adesso2014Gauss}%
  \BibitemOpen
  \bibfield  {author} {\bibinfo {author} {\bibfnamefont {G.}~\bibnamefont
  {Adesso}}, \bibinfo {author} {\bibfnamefont {S.}~\bibnamefont {Ragy}}, \ and\
  \bibinfo {author} {\bibfnamefont {A.~R.}\ \bibnamefont {Lee}},\ }\bibfield
  {title} {\emph {\bibinfo {title} {Continuous variable quantum information:
  Gaussian states and beyond},}\ }\href
  {http://dx.doi.org/10.1142/S1230161214400010} {\bibfield  {journal} {\bibinfo
   {journal} {Open Syst. Inf. Dyn.}\ }\textbf {\bibinfo {volume} {21}},\
  \bibinfo {pages} {1440001} (\bibinfo {year} {2014})}\BibitemShut {NoStop}%
\bibitem [{\citenamefont {Wakakuwa}(2017)}]{Wakakuwa2017}%
  \BibitemOpen
  \bibfield  {author} {\bibinfo {author} {\bibfnamefont {E.}~\bibnamefont
  {Wakakuwa}},\ }\href {https://arxiv.org/abs/1709.07248} {\emph {\bibinfo
  {title} {Operational resource theory of non-markovianity},}\ } (\bibinfo
  {year} {2017}),\ \Eprint {http://arxiv.org/abs/1709.07248} {arXiv:1709.07248
  [quant-ph]} \BibitemShut {NoStop}%
\bibitem [{\citenamefont {Ollivier}\ and\ \citenamefont
  {Zurek}(2001)}]{Ollivier2001}%
  \BibitemOpen
  \bibfield  {author} {\bibinfo {author} {\bibfnamefont {H.}~\bibnamefont
  {Ollivier}}\ and\ \bibinfo {author} {\bibfnamefont {W.~H.}\ \bibnamefont
  {Zurek}},\ }\bibfield  {title} {\emph {\bibinfo {title} {Quantum discord: A
  measure of the quantumness of correlations},}\ }\href
  {http://dx.doi.org/10.1103/PhysRevLett.88.017901} {\bibfield  {journal}
  {\bibinfo  {journal} {Phys. Rev. Lett.}\ }\textbf {\bibinfo {volume} {88}},\
  \bibinfo {pages} {017901} (\bibinfo {year} {2001})}\BibitemShut {NoStop}%
\bibitem [{\citenamefont {Henderson}\ and\ \citenamefont
  {Vedral}(2001)}]{Henderson2001}%
  \BibitemOpen
  \bibfield  {author} {\bibinfo {author} {\bibfnamefont {L.}~\bibnamefont
  {Henderson}}\ and\ \bibinfo {author} {\bibfnamefont {V.}~\bibnamefont
  {Vedral}},\ }\bibfield  {title} {\emph {\bibinfo {title} {Classical, quantum
  and total correlations},}\ }\href
  {http://dx.doi.org/10.1088/0305-4470/34/35/315} {\bibfield  {journal}
  {\bibinfo  {journal} {J. Phys. A}\ }\textbf {\bibinfo {volume} {34}},\
  \bibinfo {pages} {6899} (\bibinfo {year} {2001})}\BibitemShut {NoStop}%
\bibitem [{\citenamefont {Adesso}\ \emph {et~al.}(2016)\citenamefont {Adesso},
  \citenamefont {Bromley},\ and\ \citenamefont {Cianciaruso}}]{ABC2016}%
  \BibitemOpen
  \bibfield  {author} {\bibinfo {author} {\bibfnamefont {G.}~\bibnamefont
  {Adesso}}, \bibinfo {author} {\bibfnamefont {T.~R.}\ \bibnamefont {Bromley}},
  \ and\ \bibinfo {author} {\bibfnamefont {M.}~\bibnamefont {Cianciaruso}},\
  }\bibfield  {title} {\emph {\bibinfo {title} {Measures and applications of
  quantum correlations},}\ }\href
  {http://dx.doi.org/10.1088/1751-8113/49/47/473001} {\bibfield  {journal}
  {\bibinfo  {journal} {J. Phys. A: Math. Theor.}\ }\textbf {\bibinfo {volume}
  {49}},\ \bibinfo {pages} {473001} (\bibinfo {year} {2016})}\BibitemShut
  {NoStop}%
\bibitem [{\citenamefont {Bera}\ \emph {et~al.}(2017)\citenamefont {Bera},
  \citenamefont {Das}, \citenamefont {Sadhukhan}, \citenamefont {Roy},
  \citenamefont {Sen(De)},\ and\ \citenamefont {Sen}}]{Bera2017}%
  \BibitemOpen
  \bibfield  {author} {\bibinfo {author} {\bibfnamefont {A.}~\bibnamefont
  {Bera}}, \bibinfo {author} {\bibfnamefont {T.}~\bibnamefont {Das}}, \bibinfo
  {author} {\bibfnamefont {D.}~\bibnamefont {Sadhukhan}}, \bibinfo {author}
  {\bibfnamefont {S.~S.}\ \bibnamefont {Roy}}, \bibinfo {author} {\bibfnamefont
  {A.}~\bibnamefont {Sen(De)}}, \ and\ \bibinfo {author} {\bibfnamefont
  {U.}~\bibnamefont {Sen}},\ }\bibfield  {title} {\emph {\bibinfo {title}
  {Quantum discord and its allies: a review of recent progress},}\ }\href
  {http://dx.doi.org/10.1088/1361-6633/aa872f} {\bibfield  {journal} {\bibinfo
  {journal} {Rep. Prog. Phys.}\ }\textbf {\bibinfo {volume} {81}},\ \bibinfo
  {pages} {024001} (\bibinfo {year} {2017})}\BibitemShut {NoStop}%
\bibitem [{\citenamefont {Groisman}\ \emph {et~al.}(2005)\citenamefont
  {Groisman}, \citenamefont {Popescu},\ and\ \citenamefont
  {Winter}}]{Groisman2005}%
  \BibitemOpen
  \bibfield  {author} {\bibinfo {author} {\bibfnamefont {B.}~\bibnamefont
  {Groisman}}, \bibinfo {author} {\bibfnamefont {S.}~\bibnamefont {Popescu}}, \
  and\ \bibinfo {author} {\bibfnamefont {A.}~\bibnamefont {Winter}},\
  }\bibfield  {title} {\emph {\bibinfo {title} {Quantum, classical, and total
  amount of correlations in a quantum state},}\ }\href
  {http://dx.doi.org/10.1103/PhysRevA.72.032317} {\bibfield  {journal}
  {\bibinfo  {journal} {Phys. Rev. A}\ }\textbf {\bibinfo {volume} {72}},\
  \bibinfo {pages} {032317} (\bibinfo {year} {2005})}\BibitemShut {NoStop}%
\bibitem [{\citenamefont {Anshu}\ \emph {et~al.}(2018)\citenamefont {Anshu},
  \citenamefont {Hsieh},\ and\ \citenamefont {Jain}}]{Anshu2018}%
  \BibitemOpen
  \bibfield  {author} {\bibinfo {author} {\bibfnamefont {A.}~\bibnamefont
  {Anshu}}, \bibinfo {author} {\bibfnamefont {M.-H.}\ \bibnamefont {Hsieh}}, \
  and\ \bibinfo {author} {\bibfnamefont {R.}~\bibnamefont {Jain}},\ }\bibfield
  {title} {\emph {\bibinfo {title} {Quantifying resources in general resource
  theory with catalysts},}\ }\href
  {http://dx.doi.org/10.1103/PhysRevLett.121.190504} {\bibfield  {journal}
  {\bibinfo  {journal} {Phys. Rev. Lett.}\ }\textbf {\bibinfo {volume} {121}},\
  \bibinfo {pages} {190504} (\bibinfo {year} {2018})}\BibitemShut {NoStop}%
\bibitem [{\citenamefont {Kuroiwa}\ and\ \citenamefont
  {Yamasaki}(2021)}]{PhysRevA.104.L020401}%
  \BibitemOpen
  \bibfield  {author} {\bibinfo {author} {\bibfnamefont {K.}~\bibnamefont
  {Kuroiwa}}\ and\ \bibinfo {author} {\bibfnamefont {H.}~\bibnamefont
  {Yamasaki}},\ }\bibfield  {title} {\emph {\bibinfo {title} {Asymptotically
  consistent measures of general quantum resources: Discord, non-markovianity,
  and non-gaussianity},}\ }\href
  {http://dx.doi.org/10.1103/PhysRevA.104.L020401} {\bibfield  {journal}
  {\bibinfo  {journal} {Phys. Rev. A}\ }\textbf {\bibinfo {volume} {104}},\
  \bibinfo {pages} {L020401} (\bibinfo {year} {2021})}\BibitemShut {NoStop}%
\bibitem [{\citenamefont {Horodecki}\ and\ \citenamefont
  {Oppenheim}(2013)}]{Horodecki2013b}%
  \BibitemOpen
  \bibfield  {author} {\bibinfo {author} {\bibfnamefont {M.}~\bibnamefont
  {Horodecki}}\ and\ \bibinfo {author} {\bibfnamefont {J.}~\bibnamefont
  {Oppenheim}},\ }\bibfield  {title} {\emph {\bibinfo {title} {{(Quantumness in
  the context of) resource theories}},}\ }\href
  {http://dx.doi.org/10.1142/S0217979213450197} {\bibfield  {journal} {\bibinfo
   {journal} {Int. J. Mod. Phys. B}\ }\textbf {\bibinfo {volume} {27}},\
  \bibinfo {pages} {1345019} (\bibinfo {year} {2013})}\BibitemShut {NoStop}%
\bibitem [{\citenamefont {Liu}\ \emph {et~al.}(2017)\citenamefont {Liu},
  \citenamefont {Hu},\ and\ \citenamefont {Lloyd}}]{Liu2017}%
  \BibitemOpen
  \bibfield  {author} {\bibinfo {author} {\bibfnamefont {Z.-W.}\ \bibnamefont
  {Liu}}, \bibinfo {author} {\bibfnamefont {X.}~\bibnamefont {Hu}}, \ and\
  \bibinfo {author} {\bibfnamefont {S.}~\bibnamefont {Lloyd}},\ }\bibfield
  {title} {\emph {\bibinfo {title} {Resource destroying maps},}\ }\href
  {http://dx.doi.org/10.1103/PhysRevLett.118.060502} {\bibfield  {journal}
  {\bibinfo  {journal} {Phys. Rev. Lett.}\ }\textbf {\bibinfo {volume} {118}},\
  \bibinfo {pages} {060502} (\bibinfo {year} {2017})}\BibitemShut {NoStop}%
\bibitem [{\citenamefont {Regula}(2017)}]{Regula2017}%
  \BibitemOpen
  \bibfield  {author} {\bibinfo {author} {\bibfnamefont {B.}~\bibnamefont
  {Regula}},\ }\bibfield  {title} {\emph {\bibinfo {title} {Convex geometry of
  quantum resource quantification},}\ }\href
  {http://dx.doi.org/10.1088/1751-8121/aa9100} {\bibfield  {journal} {\bibinfo
  {journal} {J. Phys. A}\ }\textbf {\bibinfo {volume} {51}},\ \bibinfo {pages}
  {045303} (\bibinfo {year} {2017})}\BibitemShut {NoStop}%
\bibitem [{\citenamefont {Bromley}\ \emph {et~al.}(2018)\citenamefont
  {Bromley}, \citenamefont {Cianciaruso}, \citenamefont {Vourekas},
  \citenamefont {Regula},\ and\ \citenamefont {Adesso}}]{Bromley2018}%
  \BibitemOpen
  \bibfield  {author} {\bibinfo {author} {\bibfnamefont {T.~R.}\ \bibnamefont
  {Bromley}}, \bibinfo {author} {\bibfnamefont {M.}~\bibnamefont
  {Cianciaruso}}, \bibinfo {author} {\bibfnamefont {S.}~\bibnamefont
  {Vourekas}}, \bibinfo {author} {\bibfnamefont {B.}~\bibnamefont {Regula}}, \
  and\ \bibinfo {author} {\bibfnamefont {G.}~\bibnamefont {Adesso}},\
  }\bibfield  {title} {\emph {\bibinfo {title} {Accessible bounds for general
  quantum resources},}\ }\href {http://dx.doi.org/10.1088/1751-8121/aacb4a}
  {\bibfield  {journal} {\bibinfo  {journal} {J. Phys. A}\ }\textbf {\bibinfo
  {volume} {51}},\ \bibinfo {pages} {325303} (\bibinfo {year}
  {2018})}\BibitemShut {NoStop}%
\bibitem [{\citenamefont {Vijayan}\ \emph {et~al.}(2020)\citenamefont
  {Vijayan}, \citenamefont {Chitambar},\ and\ \citenamefont
  {Hsieh}}]{Vijayan2019}%
  \BibitemOpen
  \bibfield  {author} {\bibinfo {author} {\bibfnamefont {M.~K.}\ \bibnamefont
  {Vijayan}}, \bibinfo {author} {\bibfnamefont {E.}~\bibnamefont {Chitambar}},
  \ and\ \bibinfo {author} {\bibfnamefont {M.-H.}\ \bibnamefont {Hsieh}},\
  }\bibfield  {title} {\emph {\bibinfo {title} {Simple bounds for one-shot
  pure-state distillation in general resource theories},}\ }\href
  {http://dx.doi.org/10.1103/PhysRevA.102.052403} {\bibfield  {journal}
  {\bibinfo  {journal} {Phys. Rev. A}\ }\textbf {\bibinfo {volume} {102}},\
  \bibinfo {pages} {052403} (\bibinfo {year} {2020})}\BibitemShut {NoStop}%
\bibitem [{\citenamefont {Gonda}\ and\ \citenamefont
  {Spekkens}(2023)}]{gonda2019monotones}%
  \BibitemOpen
  \bibfield  {author} {\bibinfo {author} {\bibfnamefont {T.}~\bibnamefont
  {Gonda}}\ and\ \bibinfo {author} {\bibfnamefont {R.~W.}\ \bibnamefont
  {Spekkens}},\ }\bibfield  {title} {\emph {\bibinfo {title} {Monotones in
  {G}eneral {R}esource {T}heories},}\ }\href
  {http://dx.doi.org/10.32408/compositionality-5-7} {\bibfield  {journal}
  {\bibinfo  {journal} {{Compositionality}}\ }\textbf {\bibinfo {volume} {5}},\
  \bibinfo {pages} {7} (\bibinfo {year} {2023})}\BibitemShut {NoStop}%
\bibitem [{\citenamefont {Fang}\ and\ \citenamefont {Liu}(2020)}]{Fang2020}%
  \BibitemOpen
  \bibfield  {author} {\bibinfo {author} {\bibfnamefont {K.}~\bibnamefont
  {Fang}}\ and\ \bibinfo {author} {\bibfnamefont {Z.-W.}\ \bibnamefont {Liu}},\
  }\bibfield  {title} {\emph {\bibinfo {title} {No-go theorems for quantum
  resource purification},}\ }\href
  {http://dx.doi.org/10.1103/PhysRevLett.125.060405} {\bibfield  {journal}
  {\bibinfo  {journal} {Phys. Rev. Lett.}\ }\textbf {\bibinfo {volume} {125}},\
  \bibinfo {pages} {060405} (\bibinfo {year} {2020})}\BibitemShut {NoStop}%
\bibitem [{\citenamefont {Regula}\ \emph {et~al.}(2020)\citenamefont {Regula},
  \citenamefont {Bu}, \citenamefont {Takagi},\ and\ \citenamefont
  {Liu}}]{Regula2020}%
  \BibitemOpen
  \bibfield  {author} {\bibinfo {author} {\bibfnamefont {B.}~\bibnamefont
  {Regula}}, \bibinfo {author} {\bibfnamefont {K.}~\bibnamefont {Bu}}, \bibinfo
  {author} {\bibfnamefont {R.}~\bibnamefont {Takagi}}, \ and\ \bibinfo {author}
  {\bibfnamefont {Z.-W.}\ \bibnamefont {Liu}},\ }\bibfield  {title} {\emph
  {\bibinfo {title} {Benchmarking one-shot distillation in general quantum
  resource theories},}\ }\href {http://dx.doi.org/10.1103/PhysRevA.101.062315}
  {\bibfield  {journal} {\bibinfo  {journal} {Phys. Rev. A}\ }\textbf {\bibinfo
  {volume} {101}},\ \bibinfo {pages} {062315} (\bibinfo {year}
  {2020})}\BibitemShut {NoStop}%
\bibitem [{\citenamefont {Sparaciari}\ \emph {et~al.}(2020)\citenamefont
  {Sparaciari}, \citenamefont {del Rio}, \citenamefont {Scandolo},
  \citenamefont {Faist},\ and\ \citenamefont {Oppenheim}}]{Sparaciari2020}%
  \BibitemOpen
  \bibfield  {author} {\bibinfo {author} {\bibfnamefont {C.}~\bibnamefont
  {Sparaciari}}, \bibinfo {author} {\bibfnamefont {L.}~\bibnamefont {del Rio}},
  \bibinfo {author} {\bibfnamefont {C.~M.}\ \bibnamefont {Scandolo}}, \bibinfo
  {author} {\bibfnamefont {P.}~\bibnamefont {Faist}}, \ and\ \bibinfo {author}
  {\bibfnamefont {J.}~\bibnamefont {Oppenheim}},\ }\bibfield  {title} {\emph
  {\bibinfo {title} {The first law of general quantum resource theories},}\
  }\href {http://dx.doi.org/10.22331/q-2020-04-30-259} {\bibfield  {journal}
  {\bibinfo  {journal} {{Quantum}}\ }\textbf {\bibinfo {volume} {4}},\ \bibinfo
  {pages} {259} (\bibinfo {year} {2020})}\BibitemShut {NoStop}%
\bibitem [{\citenamefont {Designolle}\ \emph {et~al.}(2021)\citenamefont
  {Designolle}, \citenamefont {Uola}, \citenamefont {Luoma},\ and\
  \citenamefont {Brunner}}]{Brunner2021}%
  \BibitemOpen
  \bibfield  {author} {\bibinfo {author} {\bibfnamefont {S.}~\bibnamefont
  {Designolle}}, \bibinfo {author} {\bibfnamefont {R.}~\bibnamefont {Uola}},
  \bibinfo {author} {\bibfnamefont {K.}~\bibnamefont {Luoma}}, \ and\ \bibinfo
  {author} {\bibfnamefont {N.}~\bibnamefont {Brunner}},\ }\bibfield  {title}
  {\emph {\bibinfo {title} {Set coherence: Basis-independent quantification of
  quantum coherence},}\ }\href
  {http://dx.doi.org/10.1103/PhysRevLett.126.220404} {\bibfield  {journal}
  {\bibinfo  {journal} {Phys. Rev. Lett.}\ }\textbf {\bibinfo {volume} {126}},\
  \bibinfo {pages} {220404} (\bibinfo {year} {2021})}\BibitemShut {NoStop}%
\bibitem [{\citenamefont {Regula}\ and\ \citenamefont
  {Takagi}(2021{\natexlab{a}})}]{Regula2021_fundamentallimitation}%
  \BibitemOpen
  \bibfield  {author} {\bibinfo {author} {\bibfnamefont {B.}~\bibnamefont
  {Regula}}\ and\ \bibinfo {author} {\bibfnamefont {R.}~\bibnamefont
  {Takagi}},\ }\bibfield  {title} {\emph {\bibinfo {title} {Fundamental
  limitations on distillation of quantum channel resources},}\ }\href
  {http://dx.doi.org/10.1038/s41467-021-24699-0} {\bibfield  {journal}
  {\bibinfo  {journal} {Nat. Commun.}\ }\textbf {\bibinfo {volume} {12}},\
  \bibinfo {pages} {4411} (\bibinfo {year} {2021}{\natexlab{a}})}\BibitemShut
  {NoStop}%
\bibitem [{\citenamefont {Regula}\ and\ \citenamefont
  {Takagi}(2021{\natexlab{b}})}]{Regula2021one-shot}%
  \BibitemOpen
  \bibfield  {author} {\bibinfo {author} {\bibfnamefont {B.}~\bibnamefont
  {Regula}}\ and\ \bibinfo {author} {\bibfnamefont {R.}~\bibnamefont
  {Takagi}},\ }\bibfield  {title} {\emph {\bibinfo {title} {One-shot
  manipulation of dynamical quantum resources},}\ }\href
  {http://dx.doi.org/10.1103/PhysRevLett.127.060402} {\bibfield  {journal}
  {\bibinfo  {journal} {Phys. Rev. Lett.}\ }\textbf {\bibinfo {volume} {127}},\
  \bibinfo {pages} {060402} (\bibinfo {year} {2021}{\natexlab{b}})}\BibitemShut
  {NoStop}%
\bibitem [{\citenamefont {Regula}(2022{\natexlab{a}})}]{Regula2022}%
  \BibitemOpen
  \bibfield  {author} {\bibinfo {author} {\bibfnamefont {B.}~\bibnamefont
  {Regula}},\ }\bibfield  {title} {\emph {\bibinfo {title} {Probabilistic
  transformations of quantum resources},}\ }\href
  {http://dx.doi.org/10.1103/PhysRevLett.128.110505} {\bibfield  {journal}
  {\bibinfo  {journal} {Phys. Rev. Lett.}\ }\textbf {\bibinfo {volume} {128}},\
  \bibinfo {pages} {110505} (\bibinfo {year} {2022}{\natexlab{a}})}\BibitemShut
  {NoStop}%
\bibitem [{\citenamefont {Fang}\ and\ \citenamefont {Liu}(2022)}]{Fang2022}%
  \BibitemOpen
  \bibfield  {author} {\bibinfo {author} {\bibfnamefont {K.}~\bibnamefont
  {Fang}}\ and\ \bibinfo {author} {\bibfnamefont {Z.-W.}\ \bibnamefont {Liu}},\
  }\bibfield  {title} {\emph {\bibinfo {title} {No-go theorems for quantum
  resource purification: New approach and channel theory},}\ }\href
  {http://dx.doi.org/10.1103/PRXQuantum.3.010337} {\bibfield  {journal}
  {\bibinfo  {journal} {PRX Quantum}\ }\textbf {\bibinfo {volume} {3}},\
  \bibinfo {pages} {010337} (\bibinfo {year} {2022})}\BibitemShut {NoStop}%
\bibitem [{\citenamefont
  {Regula}(2022{\natexlab{b}})}]{Regula2022tightconstraints}%
  \BibitemOpen
  \bibfield  {author} {\bibinfo {author} {\bibfnamefont {B.}~\bibnamefont
  {Regula}},\ }\bibfield  {title} {\emph {\bibinfo {title} {Tight constraints
  on probabilistic convertibility of quantum states},}\ }\href
  {http://dx.doi.org/10.22331/q-2022-09-22-817} {\bibfield  {journal} {\bibinfo
   {journal} {{Quantum}}\ }\textbf {\bibinfo {volume} {6}},\ \bibinfo {pages}
  {817} (\bibinfo {year} {2022}{\natexlab{b}})}\BibitemShut {NoStop}%
\bibitem [{\citenamefont {Lami}\ and\ \citenamefont {Regula}(2023)}]{Lami2023}%
  \BibitemOpen
  \bibfield  {author} {\bibinfo {author} {\bibfnamefont {L.}~\bibnamefont
  {Lami}}\ and\ \bibinfo {author} {\bibfnamefont {B.}~\bibnamefont {Regula}},\
  }\bibfield  {title} {\emph {\bibinfo {title} {No second law of entanglement
  manipulation after all},}\ }\href
  {http://dx.doi.org/10.1038/s41567-022-01873-9} {\bibfield  {journal}
  {\bibinfo  {journal} {Nat. Phys.}\ }\textbf {\bibinfo {volume} {19}},\
  \bibinfo {pages} {184} (\bibinfo {year} {2023})}\BibitemShut {NoStop}%
\bibitem [{\citenamefont {Berta}\ \emph {et~al.}(2023)\citenamefont {Berta},
  \citenamefont {Brand{\~{a}}o}, \citenamefont {Gour}, \citenamefont {Lami},
  \citenamefont {Plenio}, \citenamefont {Regula},\ and\ \citenamefont
  {Tomamichel}}]{berta2023gap}%
  \BibitemOpen
  \bibfield  {author} {\bibinfo {author} {\bibfnamefont {M.}~\bibnamefont
  {Berta}}, \bibinfo {author} {\bibfnamefont {F.~G. S.~L.}\ \bibnamefont
  {Brand{\~{a}}o}}, \bibinfo {author} {\bibfnamefont {G.}~\bibnamefont {Gour}},
  \bibinfo {author} {\bibfnamefont {L.}~\bibnamefont {Lami}}, \bibinfo {author}
  {\bibfnamefont {M.~B.}\ \bibnamefont {Plenio}}, \bibinfo {author}
  {\bibfnamefont {B.}~\bibnamefont {Regula}}, \ and\ \bibinfo {author}
  {\bibfnamefont {M.}~\bibnamefont {Tomamichel}},\ }\bibfield  {title} {\emph
  {\bibinfo {title} {On a gap in the proof of the generalised quantum {S}tein's
  lemma and its consequences for the reversibility of quantum resources},}\
  }\href {http://dx.doi.org/10.22331/q-2023-09-07-1103} {\bibfield  {journal}
  {\bibinfo  {journal} {{Quantum}}\ }\textbf {\bibinfo {volume} {7}},\ \bibinfo
  {pages} {1103} (\bibinfo {year} {2023})}\BibitemShut {NoStop}%
\bibitem [{\citenamefont {Horodecki}(2003)}]{Horodecki2003}%
  \BibitemOpen
  \bibfield  {author} {\bibinfo {author} {\bibfnamefont {P.}~\bibnamefont
  {Horodecki}},\ }\bibfield  {title} {\emph {\bibinfo {title} {From limits of
  quantum operations to multicopy entanglement witnesses and state-spectrum
  estimation},}\ }\href {http://dx.doi.org/10.1103/PhysRevA.68.052101}
  {\bibfield  {journal} {\bibinfo  {journal} {Phys. Rev. A}\ }\textbf {\bibinfo
  {volume} {68}},\ \bibinfo {pages} {052101} (\bibinfo {year}
  {2003})}\BibitemShut {NoStop}%
\bibitem [{\citenamefont {Horodecki}\ \emph {et~al.}(2009)\citenamefont
  {Horodecki}, \citenamefont {Horodecki}, \citenamefont {Horodecki},\ and\
  \citenamefont {Horodecki}}]{Horodecki2009}%
  \BibitemOpen
  \bibfield  {author} {\bibinfo {author} {\bibfnamefont {R.}~\bibnamefont
  {Horodecki}}, \bibinfo {author} {\bibfnamefont {P.}~\bibnamefont
  {Horodecki}}, \bibinfo {author} {\bibfnamefont {M.}~\bibnamefont
  {Horodecki}}, \ and\ \bibinfo {author} {\bibfnamefont {K.}~\bibnamefont
  {Horodecki}},\ }\bibfield  {title} {\emph {\bibinfo {title} {Quantum
  entanglement},}\ }\href {http://dx.doi.org/10.1103/RevModPhys.81.865}
  {\bibfield  {journal} {\bibinfo  {journal} {Rev. Mod. Phys.}\ }\textbf
  {\bibinfo {volume} {81}},\ \bibinfo {pages} {865} (\bibinfo {year}
  {2009})}\BibitemShut {NoStop}%
\bibitem [{\citenamefont {Piani}\ and\ \citenamefont
  {Watrous}(2009)}]{Piani2009}%
  \BibitemOpen
  \bibfield  {author} {\bibinfo {author} {\bibfnamefont {M.}~\bibnamefont
  {Piani}}\ and\ \bibinfo {author} {\bibfnamefont {J.}~\bibnamefont
  {Watrous}},\ }\bibfield  {title} {\emph {\bibinfo {title} {All entangled
  states are useful for channel discrimination},}\ }\href
  {http://dx.doi.org/10.1103/PhysRevLett.102.250501} {\bibfield  {journal}
  {\bibinfo  {journal} {Phys. Rev. Lett.}\ }\textbf {\bibinfo {volume} {102}},\
  \bibinfo {pages} {250501} (\bibinfo {year} {2009})}\BibitemShut {NoStop}%
\bibitem [{\citenamefont {Kuroiwa}\ \emph {et~al.}(2024)\citenamefont
  {Kuroiwa}, \citenamefont {Takagi}, \citenamefont {Adesso},\ and\
  \citenamefont {Yamasaki}}]{PRA}%
  \BibitemOpen
  \bibfield  {author} {\bibinfo {author} {\bibfnamefont {K.}~\bibnamefont
  {Kuroiwa}}, \bibinfo {author} {\bibfnamefont {R.}~\bibnamefont {Takagi}},
  \bibinfo {author} {\bibfnamefont {G.}~\bibnamefont {Adesso}}, \ and\ \bibinfo
  {author} {\bibfnamefont {H.}~\bibnamefont {Yamasaki}},\ }\bibfield  {title}
  {\emph {\bibinfo {title} {Robustness- and weight-based resource measures
  without convexity restriction: Multicopy witness and operational advantage in
  static and dynamical quantum resource theories},}\ }\href
  {http://dx.doi.org/10.1103/PhysRevA.109.042403} {\bibfield  {journal}
  {\bibinfo  {journal} {Phys. Rev. A}\ }\textbf {\bibinfo {volume} {109}},\
  \bibinfo {pages} {042403} (\bibinfo {year} {2024})}\BibitemShut {NoStop}%
\bibitem [{\citenamefont {Gour}\ and\ \citenamefont
  {Winter}(2019)}]{Gour2019a}%
  \BibitemOpen
  \bibfield  {author} {\bibinfo {author} {\bibfnamefont {G.}~\bibnamefont
  {Gour}}\ and\ \bibinfo {author} {\bibfnamefont {A.}~\bibnamefont {Winter}},\
  }\bibfield  {title} {\emph {\bibinfo {title} {How to quantify a dynamical
  quantum resource},}\ }\href
  {http://dx.doi.org/10.1103/PhysRevLett.123.150401} {\bibfield  {journal}
  {\bibinfo  {journal} {Phys. Rev. Lett.}\ }\textbf {\bibinfo {volume} {123}},\
  \bibinfo {pages} {150401} (\bibinfo {year} {2019})}\BibitemShut {NoStop}%
\bibitem [{\citenamefont {Liu}\ and\ \citenamefont
  {Winter}(2019)}]{Liu_ZW2019b}%
  \BibitemOpen
  \bibfield  {author} {\bibinfo {author} {\bibfnamefont {Z.-W.}\ \bibnamefont
  {Liu}}\ and\ \bibinfo {author} {\bibfnamefont {A.}~\bibnamefont {Winter}},\
  }\bibfield  {title} {\emph {\bibinfo {title} {Resource theories of quantum
  channels and the universal role of resource erasure},}\ }\href
  {https://arxiv.org/abs/1904.04201} {\bibfield  {journal} {\bibinfo  {journal}
  {arXiv preprint \href{https://arxiv.org/abs/1904.04201}{arXiv:1904.04201
  [quant-ph]}}\ } (\bibinfo {year} {2019})},\ \Eprint
  {http://arxiv.org/abs/1904.04201} {arXiv:1904.04201 [quant-ph]} \BibitemShut
  {NoStop}%
\bibitem [{\citenamefont {Liu}\ and\ \citenamefont {Yuan}(2020)}]{Liu_YC2020}%
  \BibitemOpen
  \bibfield  {author} {\bibinfo {author} {\bibfnamefont {Y.}~\bibnamefont
  {Liu}}\ and\ \bibinfo {author} {\bibfnamefont {X.}~\bibnamefont {Yuan}},\
  }\bibfield  {title} {\emph {\bibinfo {title} {Operational resource theory of
  quantum channels},}\ }\href
  {http://dx.doi.org/10.1103/PhysRevResearch.2.012035} {\bibfield  {journal}
  {\bibinfo  {journal} {Phys. Rev. Research}\ }\textbf {\bibinfo {volume}
  {2}},\ \bibinfo {pages} {012035} (\bibinfo {year} {2020})}\BibitemShut
  {NoStop}%
\bibitem [{\citenamefont {Li}\ \emph {et~al.}(2020)\citenamefont {Li},
  \citenamefont {Bu},\ and\ \citenamefont {Liu}}]{Li2018}%
  \BibitemOpen
  \bibfield  {author} {\bibinfo {author} {\bibfnamefont {L.}~\bibnamefont
  {Li}}, \bibinfo {author} {\bibfnamefont {K.}~\bibnamefont {Bu}}, \ and\
  \bibinfo {author} {\bibfnamefont {Z.-W.}\ \bibnamefont {Liu}},\ }\bibfield
  {title} {\emph {\bibinfo {title} {Quantifying the resource content of quantum
  channels: An operational approach},}\ }\href
  {http://dx.doi.org/10.1103/PhysRevA.101.022335} {\bibfield  {journal}
  {\bibinfo  {journal} {Phys. Rev. A}\ }\textbf {\bibinfo {volume} {101}},\
  \bibinfo {pages} {022335} (\bibinfo {year} {2020})}\BibitemShut {NoStop}%
\bibitem [{\citenamefont {Gour}\ and\ \citenamefont
  {Wilde}(2021)}]{Gour_Wilde2018}%
  \BibitemOpen
  \bibfield  {author} {\bibinfo {author} {\bibfnamefont {G.}~\bibnamefont
  {Gour}}\ and\ \bibinfo {author} {\bibfnamefont {M.~M.}\ \bibnamefont
  {Wilde}},\ }\bibfield  {title} {\emph {\bibinfo {title} {Entropy of a quantum
  channel},}\ }\href {http://dx.doi.org/10.1103/PhysRevResearch.3.023096}
  {\bibfield  {journal} {\bibinfo  {journal} {Phys. Rev. Research}\ }\textbf
  {\bibinfo {volume} {3}},\ \bibinfo {pages} {023096} (\bibinfo {year}
  {2021})}\BibitemShut {NoStop}%
\bibitem [{\citenamefont {Takagi}\ \emph {et~al.}(2020)\citenamefont {Takagi},
  \citenamefont {Wang},\ and\ \citenamefont {Hayashi}}]{Takagi2020}%
  \BibitemOpen
  \bibfield  {author} {\bibinfo {author} {\bibfnamefont {R.}~\bibnamefont
  {Takagi}}, \bibinfo {author} {\bibfnamefont {K.}~\bibnamefont {Wang}}, \ and\
  \bibinfo {author} {\bibfnamefont {M.}~\bibnamefont {Hayashi}},\ }\bibfield
  {title} {\emph {\bibinfo {title} {Application of the resource theory of
  channels to communication scenarios},}\ }\href
  {http://dx.doi.org/10.1103/PhysRevLett.124.120502} {\bibfield  {journal}
  {\bibinfo  {journal} {Phys. Rev. Lett.}\ }\textbf {\bibinfo {volume} {124}},\
  \bibinfo {pages} {120502} (\bibinfo {year} {2020})}\BibitemShut {NoStop}%
\bibitem [{\citenamefont {Gour}\ and\ \citenamefont
  {Scandolo}(2020)}]{gour2020dynamical}%
  \BibitemOpen
  \bibfield  {author} {\bibinfo {author} {\bibfnamefont {G.}~\bibnamefont
  {Gour}}\ and\ \bibinfo {author} {\bibfnamefont {C.~M.}\ \bibnamefont
  {Scandolo}},\ }\href@noop {} {\emph {\bibinfo {title} {Dynamical
  resources},}\ } (\bibinfo {year} {2020}),\ \Eprint
  {http://arxiv.org/abs/2101.01552} {arXiv:2101.01552 [quant-ph]} \BibitemShut
  {NoStop}%
\bibitem [{\citenamefont {Yuan}\ \emph {et~al.}(2020)\citenamefont {Yuan},
  \citenamefont {Zeng}, \citenamefont {Gao},\ and\ \citenamefont
  {Zhao}}]{yuan2020oneshot}%
  \BibitemOpen
  \bibfield  {author} {\bibinfo {author} {\bibfnamefont {X.}~\bibnamefont
  {Yuan}}, \bibinfo {author} {\bibfnamefont {P.}~\bibnamefont {Zeng}}, \bibinfo
  {author} {\bibfnamefont {M.}~\bibnamefont {Gao}}, \ and\ \bibinfo {author}
  {\bibfnamefont {Q.}~\bibnamefont {Zhao}},\ }\href@noop {} {\emph {\bibinfo
  {title} {One-shot dynamical resource theory},}\ } (\bibinfo {year} {2020}),\
  \Eprint {http://arxiv.org/abs/2012.02781} {arXiv:2012.02781 [quant-ph]}
  \BibitemShut {NoStop}%
\bibitem [{\citenamefont {Heinosaari}\ \emph {et~al.}(2015)\citenamefont
  {Heinosaari}, \citenamefont {Kiukas},\ and\ \citenamefont
  {Reitzner}}]{Heinosaari2015_NoiseRobustness}%
  \BibitemOpen
  \bibfield  {author} {\bibinfo {author} {\bibfnamefont {T.}~\bibnamefont
  {Heinosaari}}, \bibinfo {author} {\bibfnamefont {J.}~\bibnamefont {Kiukas}},
  \ and\ \bibinfo {author} {\bibfnamefont {D.}~\bibnamefont {Reitzner}},\
  }\bibfield  {title} {\emph {\bibinfo {title} {Noise robustness of the
  incompatibility of quantum measurements},}\ }\href
  {http://dx.doi.org/10.1103/PhysRevA.92.022115} {\bibfield  {journal}
  {\bibinfo  {journal} {Phys. Rev. A}\ }\textbf {\bibinfo {volume} {92}},\
  \bibinfo {pages} {022115} (\bibinfo {year} {2015})}\BibitemShut {NoStop}%
\bibitem [{\citenamefont
  {Haapasalo}(2015)}]{Haapasalo2015_RobustnessIncompatibility}%
  \BibitemOpen
  \bibfield  {author} {\bibinfo {author} {\bibfnamefont {E.}~\bibnamefont
  {Haapasalo}},\ }\bibfield  {title} {\emph {\bibinfo {title} {Robustness of
  incompatibility for quantum devices},}\ }\href
  {http://dx.doi.org/10.1088/1751-8113/48/25/255303} {\bibfield  {journal}
  {\bibinfo  {journal} {J. Phys. A: Math. Theor.}\ }\textbf {\bibinfo {volume}
  {48}},\ \bibinfo {pages} {255303} (\bibinfo {year} {2015})}\BibitemShut
  {NoStop}%
\bibitem [{\citenamefont {Guerini}\ \emph {et~al.}(2017)\citenamefont
  {Guerini}, \citenamefont {Bavaresco}, \citenamefont {Terra~Cunha},\ and\
  \citenamefont {Acín}}]{Guerini2017_MeasurmentSimulability}%
  \BibitemOpen
  \bibfield  {author} {\bibinfo {author} {\bibfnamefont {L.}~\bibnamefont
  {Guerini}}, \bibinfo {author} {\bibfnamefont {J.}~\bibnamefont {Bavaresco}},
  \bibinfo {author} {\bibfnamefont {M.}~\bibnamefont {Terra~Cunha}}, \ and\
  \bibinfo {author} {\bibfnamefont {A.}~\bibnamefont {Acín}},\ }\bibfield
  {title} {\emph {\bibinfo {title} {{Operational framework for quantum
  measurement simulability}},}\ }\href {http://dx.doi.org/10.1063/1.4994303}
  {\bibfield  {journal} {\bibinfo  {journal} {J. Math. Phys.}\ }\textbf
  {\bibinfo {volume} {58}},\ \bibinfo {pages} {092102} (\bibinfo {year}
  {2017})}\BibitemShut {NoStop}%
\bibitem [{\citenamefont {Skrzypczyk}\ and\ \citenamefont
  {Linden}(2019)}]{Skrzypczyk2019_RobustnessMeasurement}%
  \BibitemOpen
  \bibfield  {author} {\bibinfo {author} {\bibfnamefont {P.}~\bibnamefont
  {Skrzypczyk}}\ and\ \bibinfo {author} {\bibfnamefont {N.}~\bibnamefont
  {Linden}},\ }\bibfield  {title} {\emph {\bibinfo {title} {Robustness of
  measurement, discrimination games, and accessible information},}\ }\href
  {http://dx.doi.org/10.1103/PhysRevLett.122.140403} {\bibfield  {journal}
  {\bibinfo  {journal} {Phys. Rev. Lett.}\ }\textbf {\bibinfo {volume} {122}},\
  \bibinfo {pages} {140403} (\bibinfo {year} {2019})}\BibitemShut {NoStop}%
\bibitem [{\citenamefont {Skrzypczyk}\ \emph {et~al.}(2019)\citenamefont
  {Skrzypczyk}, \citenamefont {\ifmmode \check{S}\else
  \v{S}\fi{}upi\ifmmode~\acute{c}\else \'{c}\fi{}},\ and\ \citenamefont
  {Cavalcanti}}]{Skrzypczyk2019_IncompatibleMeasurements}%
  \BibitemOpen
  \bibfield  {author} {\bibinfo {author} {\bibfnamefont {P.}~\bibnamefont
  {Skrzypczyk}}, \bibinfo {author} {\bibfnamefont {I.}~\bibnamefont {\ifmmode
  \check{S}\else \v{S}\fi{}upi\ifmmode~\acute{c}\else \'{c}\fi{}}}, \ and\
  \bibinfo {author} {\bibfnamefont {D.}~\bibnamefont {Cavalcanti}},\ }\bibfield
   {title} {\emph {\bibinfo {title} {All sets of incompatible measurements give
  an advantage in quantum state discrimination},}\ }\href
  {http://dx.doi.org/10.1103/PhysRevLett.122.130403} {\bibfield  {journal}
  {\bibinfo  {journal} {Phys. Rev. Lett.}\ }\textbf {\bibinfo {volume} {122}},\
  \bibinfo {pages} {130403} (\bibinfo {year} {2019})}\BibitemShut {NoStop}%
\bibitem [{\citenamefont {Oszmaniec}\ and\ \citenamefont
  {Biswas}(2019)}]{Oszmaniec2019operational}%
  \BibitemOpen
  \bibfield  {author} {\bibinfo {author} {\bibfnamefont {M.}~\bibnamefont
  {Oszmaniec}}\ and\ \bibinfo {author} {\bibfnamefont {T.}~\bibnamefont
  {Biswas}},\ }\bibfield  {title} {\emph {\bibinfo {title} {Operational
  relevance of resource theories of quantum measurements},}\ }\href
  {http://dx.doi.org/10.22331/q-2019-04-26-133} {\bibfield  {journal} {\bibinfo
   {journal} {{Quantum}}\ }\textbf {\bibinfo {volume} {3}},\ \bibinfo {pages}
  {133} (\bibinfo {year} {2019})}\BibitemShut {NoStop}%
\bibitem [{\citenamefont {Guff}\ \emph {et~al.}(2021)\citenamefont {Guff},
  \citenamefont {McMahon}, \citenamefont {Sanders},\ and\ \citenamefont
  {Gilchrist}}]{Guff2021_ResourceMeasurement}%
  \BibitemOpen
  \bibfield  {author} {\bibinfo {author} {\bibfnamefont {T.}~\bibnamefont
  {Guff}}, \bibinfo {author} {\bibfnamefont {N.~A.}\ \bibnamefont {McMahon}},
  \bibinfo {author} {\bibfnamefont {Y.~R.}\ \bibnamefont {Sanders}}, \ and\
  \bibinfo {author} {\bibfnamefont {A.}~\bibnamefont {Gilchrist}},\ }\bibfield
  {title} {\emph {\bibinfo {title} {A resource theory of quantum
  measurements},}\ }\href {http://dx.doi.org/10.1088/1751-8121/abed67}
  {\bibfield  {journal} {\bibinfo  {journal} {J. Phys. A: Math. Theor.}\
  }\textbf {\bibinfo {volume} {54}},\ \bibinfo {pages} {225301} (\bibinfo
  {year} {2021})}\BibitemShut {NoStop}%
\bibitem [{\citenamefont {Lewenstein}\ and\ \citenamefont
  {Sanpera}(1998)}]{Lewenstein1998}%
  \BibitemOpen
  \bibfield  {author} {\bibinfo {author} {\bibfnamefont {M.}~\bibnamefont
  {Lewenstein}}\ and\ \bibinfo {author} {\bibfnamefont {A.}~\bibnamefont
  {Sanpera}},\ }\bibfield  {title} {\emph {\bibinfo {title} {Separability and
  entanglement of composite quantum systems},}\ }\href
  {http://dx.doi.org/10.1103/PhysRevLett.80.2261} {\bibfield  {journal}
  {\bibinfo  {journal} {Phys. Rev. Lett.}\ }\textbf {\bibinfo {volume} {80}},\
  \bibinfo {pages} {2261} (\bibinfo {year} {1998})}\BibitemShut {NoStop}%
\bibitem [{\citenamefont {Skrzypczyk}\ \emph {et~al.}(2014)\citenamefont
  {Skrzypczyk}, \citenamefont {Navascu\'es},\ and\ \citenamefont
  {Cavalcanti}}]{Skrzypczyk2014}%
  \BibitemOpen
  \bibfield  {author} {\bibinfo {author} {\bibfnamefont {P.}~\bibnamefont
  {Skrzypczyk}}, \bibinfo {author} {\bibfnamefont {M.}~\bibnamefont
  {Navascu\'es}}, \ and\ \bibinfo {author} {\bibfnamefont {D.}~\bibnamefont
  {Cavalcanti}},\ }\bibfield  {title} {\emph {\bibinfo {title} {Quantifying
  einstein-podolsky-rosen steering},}\ }\href
  {http://dx.doi.org/10.1103/PhysRevLett.112.180404} {\bibfield  {journal}
  {\bibinfo  {journal} {Phys. Rev. Lett.}\ }\textbf {\bibinfo {volume} {112}},\
  \bibinfo {pages} {180404} (\bibinfo {year} {2014})}\BibitemShut {NoStop}%
\bibitem [{\citenamefont {Pusey}(2015)}]{Pusey2015}%
  \BibitemOpen
  \bibfield  {author} {\bibinfo {author} {\bibfnamefont {M.~F.}\ \bibnamefont
  {Pusey}},\ }\bibfield  {title} {\emph {\bibinfo {title} {Verifying the
  quantumness of a channel with an untrusted device},}\ }\href
  {http://dx.doi.org/10.1364/JOSAB.32.000A56} {\bibfield  {journal} {\bibinfo
  {journal} {J. Opt. Soc. Am. B}\ }\textbf {\bibinfo {volume} {32}},\ \bibinfo
  {pages} {A56} (\bibinfo {year} {2015})}\BibitemShut {NoStop}%
\bibitem [{\citenamefont {Cavalcanti}\ and\ \citenamefont
  {Skrzypczyk}(2016)}]{Cavalcanti2016}%
  \BibitemOpen
  \bibfield  {author} {\bibinfo {author} {\bibfnamefont {D.}~\bibnamefont
  {Cavalcanti}}\ and\ \bibinfo {author} {\bibfnamefont {P.}~\bibnamefont
  {Skrzypczyk}},\ }\bibfield  {title} {\emph {\bibinfo {title} {Quantitative
  relations between measurement incompatibility, quantum steering, and
  nonlocality},}\ }\href {http://dx.doi.org/10.1103/PhysRevA.93.052112}
  {\bibfield  {journal} {\bibinfo  {journal} {Phys. Rev. A}\ }\textbf {\bibinfo
  {volume} {93}},\ \bibinfo {pages} {052112} (\bibinfo {year}
  {2016})}\BibitemShut {NoStop}%
\bibitem [{\citenamefont {Bu}\ \emph {et~al.}(2018)\citenamefont {Bu},
  \citenamefont {Anand},\ and\ \citenamefont {Singh}}]{Kaifeng2018}%
  \BibitemOpen
  \bibfield  {author} {\bibinfo {author} {\bibfnamefont {K.}~\bibnamefont
  {Bu}}, \bibinfo {author} {\bibfnamefont {N.}~\bibnamefont {Anand}}, \ and\
  \bibinfo {author} {\bibfnamefont {U.}~\bibnamefont {Singh}},\ }\bibfield
  {title} {\emph {\bibinfo {title} {Asymmetry and coherence weight of quantum
  states},}\ }\href {http://dx.doi.org/10.1103/PhysRevA.97.032342} {\bibfield
  {journal} {\bibinfo  {journal} {Phys. Rev. A}\ }\textbf {\bibinfo {volume}
  {97}},\ \bibinfo {pages} {032342} (\bibinfo {year} {2018})}\BibitemShut
  {NoStop}%
\bibitem [{\citenamefont {Ducuara}\ and\ \citenamefont
  {Skrzypczyk}(2020)}]{Ducuara2020}%
  \BibitemOpen
  \bibfield  {author} {\bibinfo {author} {\bibfnamefont {A.~F.}\ \bibnamefont
  {Ducuara}}\ and\ \bibinfo {author} {\bibfnamefont {P.}~\bibnamefont
  {Skrzypczyk}},\ }\bibfield  {title} {\emph {\bibinfo {title} {Operational
  interpretation of weight-based resource quantifiers in convex quantum
  resource theories},}\ }\href
  {http://dx.doi.org/10.1103/PhysRevLett.125.110401} {\bibfield  {journal}
  {\bibinfo  {journal} {Phys. Rev. Lett.}\ }\textbf {\bibinfo {volume} {125}},\
  \bibinfo {pages} {110401} (\bibinfo {year} {2020})}\BibitemShut {NoStop}%
\bibitem [{\citenamefont {Uola}\ \emph {et~al.}(2020)\citenamefont {Uola},
  \citenamefont {Bullock}, \citenamefont {Kraft}, \citenamefont
  {Pellonp\"a\"a},\ and\ \citenamefont {Brunner}}]{Uola2020}%
  \BibitemOpen
  \bibfield  {author} {\bibinfo {author} {\bibfnamefont {R.}~\bibnamefont
  {Uola}}, \bibinfo {author} {\bibfnamefont {T.}~\bibnamefont {Bullock}},
  \bibinfo {author} {\bibfnamefont {T.}~\bibnamefont {Kraft}}, \bibinfo
  {author} {\bibfnamefont {J.-P.}\ \bibnamefont {Pellonp\"a\"a}}, \ and\
  \bibinfo {author} {\bibfnamefont {N.}~\bibnamefont {Brunner}},\ }\bibfield
  {title} {\emph {\bibinfo {title} {All quantum resources provide an advantage
  in exclusion tasks},}\ }\href
  {http://dx.doi.org/10.1103/PhysRevLett.125.110402} {\bibfield  {journal}
  {\bibinfo  {journal} {Phys. Rev. Lett.}\ }\textbf {\bibinfo {volume} {125}},\
  \bibinfo {pages} {110402} (\bibinfo {year} {2020})}\BibitemShut {NoStop}%
\bibitem [{\citenamefont {Kimura}(2003)}]{KIMURA2003339}%
  \BibitemOpen
  \bibfield  {author} {\bibinfo {author} {\bibfnamefont {G.}~\bibnamefont
  {Kimura}},\ }\bibfield  {title} {\emph {\bibinfo {title} {The bloch vector
  for n-level systems},}\ }\href
  {http://dx.doi.org/https://doi.org/10.1016/S0375-9601(03)00941-1} {\bibfield
  {journal} {\bibinfo  {journal} {Phys. Lett. A}\ }\textbf {\bibinfo {volume}
  {314}},\ \bibinfo {pages} {339} (\bibinfo {year} {2003})}\BibitemShut
  {NoStop}%
\bibitem [{\citenamefont {Bertlmann}\ and\ \citenamefont
  {Krammer}(2008)}]{Bertlmann_2008}%
  \BibitemOpen
  \bibfield  {author} {\bibinfo {author} {\bibfnamefont {R.~A.}\ \bibnamefont
  {Bertlmann}}\ and\ \bibinfo {author} {\bibfnamefont {P.}~\bibnamefont
  {Krammer}},\ }\bibfield  {title} {\emph {\bibinfo {title} {Bloch vectors for
  qudits},}\ }\href {http://dx.doi.org/10.1088/1751-8113/41/23/235303}
  {\bibfield  {journal} {\bibinfo  {journal} {J. Phys. A: Math. Theor.}\
  }\textbf {\bibinfo {volume} {41}},\ \bibinfo {pages} {235303} (\bibinfo
  {year} {2008})}\BibitemShut {NoStop}%
\bibitem [{\citenamefont {Byrd}\ and\ \citenamefont
  {Khaneja}(2003)}]{Byrd2003_characterization}%
  \BibitemOpen
  \bibfield  {author} {\bibinfo {author} {\bibfnamefont {M.~S.}\ \bibnamefont
  {Byrd}}\ and\ \bibinfo {author} {\bibfnamefont {N.}~\bibnamefont {Khaneja}},\
  }\bibfield  {title} {\emph {\bibinfo {title} {Characterization of the
  positivity of the density matrix in terms of the coherence vector
  representation},}\ }\href {http://dx.doi.org/10.1103/PhysRevA.68.062322}
  {\bibfield  {journal} {\bibinfo  {journal} {Phys. Rev. A}\ }\textbf {\bibinfo
  {volume} {68}},\ \bibinfo {pages} {062322} (\bibinfo {year}
  {2003})}\BibitemShut {NoStop}%
\bibitem [{\citenamefont {Ekert}\ \emph {et~al.}(2002)\citenamefont {Ekert},
  \citenamefont {Alves}, \citenamefont {Oi}, \citenamefont {Horodecki},
  \citenamefont {Horodecki},\ and\ \citenamefont {Kwek}}]{Ekert2002}%
  \BibitemOpen
  \bibfield  {author} {\bibinfo {author} {\bibfnamefont {A.~K.}\ \bibnamefont
  {Ekert}}, \bibinfo {author} {\bibfnamefont {C.~M.}\ \bibnamefont {Alves}},
  \bibinfo {author} {\bibfnamefont {D.~K.~L.}\ \bibnamefont {Oi}}, \bibinfo
  {author} {\bibfnamefont {M.}~\bibnamefont {Horodecki}}, \bibinfo {author}
  {\bibfnamefont {P.}~\bibnamefont {Horodecki}}, \ and\ \bibinfo {author}
  {\bibfnamefont {L.~C.}\ \bibnamefont {Kwek}},\ }\bibfield  {title} {\emph
  {\bibinfo {title} {Direct estimations of linear and nonlinear functionals of
  a quantum state},}\ }\href {http://dx.doi.org/10.1103/PhysRevLett.88.217901}
  {\bibfield  {journal} {\bibinfo  {journal} {Phys. Rev. Lett.}\ }\textbf
  {\bibinfo {volume} {88}},\ \bibinfo {pages} {217901} (\bibinfo {year}
  {2002})}\BibitemShut {NoStop}%
\bibitem [{\citenamefont {Baumgratz}\ and\ \citenamefont
  {Datta}(2016)}]{Baumgratz2016}%
  \BibitemOpen
  \bibfield  {author} {\bibinfo {author} {\bibfnamefont {T.}~\bibnamefont
  {Baumgratz}}\ and\ \bibinfo {author} {\bibfnamefont {A.}~\bibnamefont
  {Datta}},\ }\bibfield  {title} {\emph {\bibinfo {title} {Quantum enhanced
  estimation of a multidimensional field},}\ }\href
  {http://dx.doi.org/10.1103/PhysRevLett.116.030801} {\bibfield  {journal}
  {\bibinfo  {journal} {Phys. Rev. Lett.}\ }\textbf {\bibinfo {volume} {116}},\
  \bibinfo {pages} {030801} (\bibinfo {year} {2016})}\BibitemShut {NoStop}%
\bibitem [{\citenamefont {Correa}\ \emph {et~al.}(2014)\citenamefont {Correa},
  \citenamefont {Palao}, \citenamefont {Alonso},\ and\ \citenamefont
  {Adesso}}]{Correa2014}%
  \BibitemOpen
  \bibfield  {author} {\bibinfo {author} {\bibfnamefont {L.~A.}\ \bibnamefont
  {Correa}}, \bibinfo {author} {\bibfnamefont {J.~P.}\ \bibnamefont {Palao}},
  \bibinfo {author} {\bibfnamefont {D.}~\bibnamefont {Alonso}}, \ and\ \bibinfo
  {author} {\bibfnamefont {G.}~\bibnamefont {Adesso}},\ }\bibfield  {title}
  {\emph {\bibinfo {title} {Quantum-enhanced absorption refrigerators},}\
  }\href {http://dx.doi.org/10.1038/srep03949} {\bibfield  {journal} {\bibinfo
  {journal} {Sci. Rep.}\ }\textbf {\bibinfo {volume} {4}},\ \bibinfo {pages}
  {3949} (\bibinfo {year} {2014})}\BibitemShut {NoStop}%
\bibitem [{\citenamefont {Wojew{\'{o}}dka-{\'{S}}ci{\k{a}}{\.{z}}ko}\ \emph
  {et~al.}(2024)\citenamefont {Wojew{\'{o}}dka-{\'{S}}ci{\k{a}}{\.{z}}ko},
  \citenamefont {Pucha{\l{}}a},\ and\ \citenamefont {Korzekwa}}]{Korzekwa2023}%
  \BibitemOpen
  \bibfield  {author} {\bibinfo {author} {\bibfnamefont {H.}~\bibnamefont
  {Wojew{\'{o}}dka-{\'{S}}ci{\k{a}}{\.{z}}ko}}, \bibinfo {author}
  {\bibfnamefont {Z.}~\bibnamefont {Pucha{\l{}}a}}, \ and\ \bibinfo {author}
  {\bibfnamefont {K.}~\bibnamefont {Korzekwa}},\ }\bibfield  {title} {\emph
  {\bibinfo {title} {Resource engines},}\ }\href
  {http://dx.doi.org/10.22331/q-2024-01-10-1222} {\bibfield  {journal}
  {\bibinfo  {journal} {{Quantum}}\ }\textbf {\bibinfo {volume} {8}},\ \bibinfo
  {pages} {1222} (\bibinfo {year} {2024})}\BibitemShut {NoStop}%
\bibitem [{\citenamefont {Yamasaki}\ \emph {et~al.}(2022)\citenamefont
  {Yamasaki}, \citenamefont {Morelli}, \citenamefont {Miethlinger},
  \citenamefont {Bavaresco}, \citenamefont {Friis},\ and\ \citenamefont
  {Huber}}]{Yamasaki2022activationofgenuine}%
  \BibitemOpen
  \bibfield  {author} {\bibinfo {author} {\bibfnamefont {H.}~\bibnamefont
  {Yamasaki}}, \bibinfo {author} {\bibfnamefont {S.}~\bibnamefont {Morelli}},
  \bibinfo {author} {\bibfnamefont {M.}~\bibnamefont {Miethlinger}}, \bibinfo
  {author} {\bibfnamefont {J.}~\bibnamefont {Bavaresco}}, \bibinfo {author}
  {\bibfnamefont {N.}~\bibnamefont {Friis}}, \ and\ \bibinfo {author}
  {\bibfnamefont {M.}~\bibnamefont {Huber}},\ }\bibfield  {title} {\emph
  {\bibinfo {title} {Activation of genuine multipartite entanglement: {B}eyond
  the single-copy paradigm of entanglement characterisation},}\ }\href
  {http://dx.doi.org/10.22331/q-2022-04-25-695} {\bibfield  {journal} {\bibinfo
   {journal} {{Quantum}}\ }\textbf {\bibinfo {volume} {6}},\ \bibinfo {pages}
  {695} (\bibinfo {year} {2022})}\BibitemShut {NoStop}%
\bibitem [{\citenamefont {Palazuelos}\ and\ \citenamefont
  {Vicente}(2022)}]{Palazuelos2022genuinemultipartite}%
  \BibitemOpen
  \bibfield  {author} {\bibinfo {author} {\bibfnamefont {C.}~\bibnamefont
  {Palazuelos}}\ and\ \bibinfo {author} {\bibfnamefont {J.~I.~d.}\ \bibnamefont
  {Vicente}},\ }\bibfield  {title} {\emph {\bibinfo {title} {Genuine
  multipartite entanglement of quantum states in the multiple-copy scenario},}\
  }\href {http://dx.doi.org/10.22331/q-2022-06-13-735} {\bibfield  {journal}
  {\bibinfo  {journal} {{Quantum}}\ }\textbf {\bibinfo {volume} {6}},\ \bibinfo
  {pages} {735} (\bibinfo {year} {2022})}\BibitemShut {NoStop}%
\bibitem [{\citenamefont {Girolami}\ \emph {et~al.}(2013)\citenamefont
  {Girolami}, \citenamefont {Tufarelli},\ and\ \citenamefont
  {Adesso}}]{Girolami2013}%
  \BibitemOpen
  \bibfield  {author} {\bibinfo {author} {\bibfnamefont {D.}~\bibnamefont
  {Girolami}}, \bibinfo {author} {\bibfnamefont {T.}~\bibnamefont {Tufarelli}},
  \ and\ \bibinfo {author} {\bibfnamefont {G.}~\bibnamefont {Adesso}},\
  }\bibfield  {title} {\emph {\bibinfo {title} {Characterizing nonclassical
  correlations via local quantum uncertainty},}\ }\href
  {http://dx.doi.org/10.1103/PhysRevLett.110.240402} {\bibfield  {journal}
  {\bibinfo  {journal} {Phys. Rev. Lett.}\ }\textbf {\bibinfo {volume} {110}},\
  \bibinfo {pages} {240402} (\bibinfo {year} {2013})}\BibitemShut {NoStop}%
\bibitem [{\citenamefont {Girolami}\ \emph {et~al.}(2014)\citenamefont
  {Girolami}, \citenamefont {Souza}, \citenamefont {Giovannetti}, \citenamefont
  {Tufarelli}, \citenamefont {Filgueiras}, \citenamefont {Sarthour},
  \citenamefont {Soares-Pinto}, \citenamefont {Oliveira},\ and\ \citenamefont
  {Adesso}}]{Girolami2014}%
  \BibitemOpen
  \bibfield  {author} {\bibinfo {author} {\bibfnamefont {D.}~\bibnamefont
  {Girolami}}, \bibinfo {author} {\bibfnamefont {A.~M.}\ \bibnamefont {Souza}},
  \bibinfo {author} {\bibfnamefont {V.}~\bibnamefont {Giovannetti}}, \bibinfo
  {author} {\bibfnamefont {T.}~\bibnamefont {Tufarelli}}, \bibinfo {author}
  {\bibfnamefont {J.~G.}\ \bibnamefont {Filgueiras}}, \bibinfo {author}
  {\bibfnamefont {R.~S.}\ \bibnamefont {Sarthour}}, \bibinfo {author}
  {\bibfnamefont {D.~O.}\ \bibnamefont {Soares-Pinto}}, \bibinfo {author}
  {\bibfnamefont {I.~S.}\ \bibnamefont {Oliveira}}, \ and\ \bibinfo {author}
  {\bibfnamefont {G.}~\bibnamefont {Adesso}},\ }\bibfield  {title} {\emph
  {\bibinfo {title} {Quantum discord determines the interferometric power of
  quantum states},}\ }\href {http://dx.doi.org/10.1103/PhysRevLett.112.210401}
  {\bibfield  {journal} {\bibinfo  {journal} {Phys. Rev. Lett.}\ }\textbf
  {\bibinfo {volume} {112}},\ \bibinfo {pages} {210401} (\bibinfo {year}
  {2014})}\BibitemShut {NoStop}%
\bibitem [{\citenamefont {Farace}\ \emph {et~al.}(2014)\citenamefont {Farace},
  \citenamefont {De~Pasquale}, \citenamefont {Rigovacca},\ and\ \citenamefont
  {Giovannetti}}]{Farace2014}%
  \BibitemOpen
  \bibfield  {author} {\bibinfo {author} {\bibfnamefont {A.}~\bibnamefont
  {Farace}}, \bibinfo {author} {\bibfnamefont {A.}~\bibnamefont {De~Pasquale}},
  \bibinfo {author} {\bibfnamefont {L.}~\bibnamefont {Rigovacca}}, \ and\
  \bibinfo {author} {\bibfnamefont {V.}~\bibnamefont {Giovannetti}},\
  }\bibfield  {title} {\emph {\bibinfo {title} {Discriminating strength: a bona
  fide measure of non-classical correlations},}\ }\href
  {http://dx.doi.org/10.1088/1367-2630/16/7/073010} {\bibfield  {journal}
  {\bibinfo  {journal} {New J. Phys.}\ }\textbf {\bibinfo {volume} {16}},\
  \bibinfo {pages} {073010} (\bibinfo {year} {2014})}\BibitemShut {NoStop}%
\end{thebibliography}%
\end{document}